\documentclass[twocolumn]{aastex631}

\DeclareSymbolFont{matha}{OML}{txmi}{m}{it}
\DeclareMathSymbol{\varv}{\mathord}{matha}{118}

\newcommand{\rhog}{\rho_g}
\newcommand{\rhop}{\rho_p}
\newcommand{\vecA}{\textit{\textbf{A}}}
\newcommand{\vecx}{\textit{\textbf{x}}}
\newcommand{\vecy}{\textit{\textbf{y}}}
\newcommand{\vecz}{\textit{\textbf{z}}}
\newcommand{\vecu}{\textit{\textbf{u}}}

\newcommand{\vecv}{{\bm{\varv}}}
\newcommand{\vecI}{\textit{\textbf{I}}}
\newcommand{\vecOmega}{\boldsymbol{\Omega}}
\newcommand{\fturb}{\mathbf{f}_{\rm{turb}}}
\newcommand{\alphaD}{\alpha_{\textrm{D}}}
\newcommand{\alphaDz}{\alpha_{\textrm{D},z}}
\newcommand{\epscrit}{\epsilon_{\rm{crit}}}
\newcommand{\epscritLY}{\epsilon_{\textrm{crit,LY21}}}
\newcommand{\Zcrit}{Z_{\rm{crit}}}
\newcommand{\Zcrithp}{Z_{\rm{crit,Gaussian} \ \it{H_p}}}
\newcommand{\Zcritalpha}{Z'_{\rm{crit}}}
\newcommand{\Hpsi}{H_{p,\rm{SI}}}
\newcommand{\Hpturb}{H_{p,\rm{\alpha}}}
\newcommand{\Hpsiturb}{H_{p,\rm{SI \_ Turb}}}
\newcommand{\tildeG}{\widetilde{G}}
\newcommand{\sigmapz}{\sigma_{p,z}}

\shorttitle{Planetesimal formation thresholds with turbulence}
\shortauthors{Lim et al.}

\graphicspath{{./}{figures/}}
\usepackage{amsmath}
\usepackage{amssymb}
\usepackage{bm}
\usepackage{comment}
\usepackage{hyperref}
\usepackage[normalem]{ulem}
\defcitealias{Li21}{LY21}
\usepackage{CJKutf8}
\newcommand{\ctext}[1]{\textup{\begin{CJK*}{UTF8}{bkai}#1\ignorespacesafterend\end{CJK*}}}

\newcommand\textkorean[1]{%
\begin{CJK}{UTF8}{mj}#1\ignorespacesafterend\end{CJK}}

\begin{document}

\title{Streaming Instability and Turbulence: Conditions for Planetesimal Formation}

\correspondingauthor{Jeonghoon Lim}
\email{jhlim@iastate.edu}

\author[0000-0003-2719-6640]{Jeonghoon Lim (\textkorean{임정훈})
}
\affiliation{Department of Physics and Astronomy, Iowa State University, Ames, IA 50010, USA}

\author[0000-0002-3771-8054]{Jacob B. Simon}
\affiliation{Department of Physics and Astronomy, Iowa State University, Ames, IA 50010, USA}

\author[0000-0001-9222-4367]{Rixin Li
\begin{CJK*}{UTF8}{gkai}(李日新)\end{CJK*}
}
\altaffiliation{51 Pegasi b Fellow}
\affiliation{Department of Astronomy, University of California, Berkeley, Berkeley, CA 94720, USA}

\author[0000-0001-5032-1396]{Philip J. Armitage}
\affiliation{Center for Computational Astrophysics, Flatiron Institute, 162 Fifth Avenue, New York, NY 10010, USA}
\affiliation{Department of Physics and Astronomy, Stony Brook University, Stony Brook, NY 11794, USA}

\author[0000-0001-6259-3575]{Daniel Carrera}
\affiliation{Department of Physics and Astronomy, Iowa State University, Ames, IA 50010, USA}

\author[0000-0002-3768-7542]{Wladimir Lyra}
\affiliation{New Mexico State University, Department of Astronomy, PO Box 30001 MSC 4500, Las Cruces, NM 88001, USA}

\author[0000-0002-5000-2747]{David G. Rea}
\affiliation{Department of Physics and Astronomy, Iowa State University, Ames, IA 50010, USA}

\author[0000-0003-2589-5034]{Chao-Chin Yang (\ctext{楊朝欽})}
\affiliation{Department of Physics and Astronomy, The University of Alabama,
    Box~870324, Tuscaloosa, AL~35487-0324, USA}

\author[0000-0002-3644-8726]{Andrew N. Youdin}
\affiliation{Department of Astronomy and Steward Observatory, University of Arizona, Tucson, Arizona 85721, USA}
\affiliation{The Lunar and Planetary Laboratory, University of Arizona, Tucson, Arizona 85721, USA}

\begin{abstract}
The streaming instability (SI) is a leading candidate for planetesimal formation, which can concentrate solids through two-way aerodynamic interactions with the gas. The resulting concentrations can become sufficiently dense to collapse under particle self-gravity, forming planetesimals. Previous studies have carried out large parameter surveys to establish the critical particle to gas surface density ratio ($Z$), above which SI-induced concentration triggers planetesimal formation. The threshold $Z$ depends on the dimensionless stopping time ($\tau_s$, a proxy for dust size). However, these studies neglected both particle self-gravity and external turbulence. Here, we perform 3D stratified shearing box simulations with both particle self-gravity and turbulent forcing, which we characterize via $\alphaD$ that measures turbulent diffusion. We find that forced turbulence, at amplitudes plausibly present in some protoplanetary disks, can increase the threshold $Z$ by up to an order of magnitude.  For example, for $\tau_s = 0.01$, planetesimal formation occurs when $Z \gtrsim 0.06$, $\gtrsim 0.1$, and $\gtrsim 0.2$ at $\alphaD = 10^{-4}$, $10^{-3.5}$, and $10^{-3}$, respectively. We provide a single fit to the critical $Z$ as a function of $\alphaD$ and $\tau_s$ required for the SI to work (though limited to the range $\tau_s = 0.01$--0.1). Our simulations also show that planetesimal formation requires a mid-plane particle-to-gas density ratio that exceeds unity, with the critical value being independent of $\alphaD$. Finally, we provide the estimation of particle scale height that accounts for both particle feedback and external turbulence. 
\end{abstract}




\section{Introduction}\label{sec:Intro}
Planet formation involves a variety of physical processes ranging from collisions of tiny dust grains to migration of massive planets within a circumstellar disk. It must be efficient and rapid; it requires that growth from micron-sized dust into fully-fledged planets of up to $10^5$ km in scale be completed within several Myr before gas dissipates, possibly even earlier as suggested by some observations (e.g., \citealt{Manara18}, \citealt{2020NatureIRS63}).  The intermediate stage of the process, where millimeter- to centimeter-sized pebbles coalesce into kilometer-scale planetesimals, entails one of the most challenging questions to answer: how do the planetesimals form out of their much smaller pebble constituents (see \citealt{SimonCometsIII} for a recent review)?
 
The challenge in answering this question lies with the growth barriers that must be overcome in order to form planetesimals. First, while collisional coagulation of dust grains is effective at producing grains of size $\sim$mm--cm, growth beyond this scale is stalled by the fragmentation and/or bouncing, the latter of which may even prevent growth even to the fragmentation limit (\citealt{Zsom10}, see also \citealt{dominik2023bouncing} for more recent work). 
of these larger grains in the inner regions ($\lesssim 10$ au) of protoplanetry disks (PPDs; e.g., \citealt{Blum08,Guttler10, Zsom10, Birnstiel12}). Second, as particles experience orbit at the Keplerian speed, they feel a headwind caused by the more slowly rotating, sub-Keplerian gas; this headwind removes angular momentum from the  particles, causing them to drift radially inward. Since the drift timescale is short compared to the disk lifetime, \citep[e.g., $\sim 300$ orbital periods for cm-sized particles at 50 AU;][]{1976PThPh..56.1756A}, the rapid radial drift imposes limits on the growth of particles in the outer regions \citep{Birnstiel12}.

These growth barriers can be bypassed by the streaming instability (SI, \citealt{YG05}), which arises from angular momentum exchange between gas and solid particles via aerodynamic coupling. Linear studies of the SI \citep{YG05, YJ07} have demonstrated that the coupled gas-solid system in disks is unstable, with the exponential growth rate depending mainly on the dimensionless stopping time of particles ($\tau_s$) and particle-to-gas density ratio.  Beyond the linear regime, numerical simulations have shown that nonlinear evolution of the SI leads to particles concentrating in narrow filaments, and that under some circumstances, these filaments can reach sufficiently high densities that gravitational collapse ensues, forming planetesimals \citep{Johansen2007Rapidformation,Johansen09b,Johansen2015SciA,Simon2016, Schafer2017, Abod2019}.

Several numerical studies have delved into the nature of the SI clumping by examining the critical ratio of pebble surface density to gas surface density ($Z$), beyond which the SI triggers strong clumping that facilitates planetesimal formation. These studies have explored large regions in ($\tau_s, Z$) parameter space (\citealt{Carrera2015, Yang2017, Li21}, hereafter \citetalias{Li21}), quantifying a boundary (the ``clumping boundary") that determines which combinations of these two parameters could result in strong clumping.  They have identified that the lowest critical $Z$ values occur around $\tau_s$ $\sim$ 0.1 - 0.3, with the exact critical $Z$ values depending on the specific numerical setup (e.g., vertical boundary conditions and domain size; see \citealt{Li21}, Section 4.1.4). While thresholds for the SI clumping have been established for much of parameter space, there is still work to be done to understand this boundary under the presence of more realistic physics, e.g., including particle self-gravity and external turbulence.

First, the previous SI simulations did not take (driven) external turbulence\footnote{Turbulence is driven within the settled particle layers by the SI \citep{Johansen2007SIturb, YangZhu2021MNRAS} or other instabilities when SI is inactive \citep{Sengupta_Umurhan23} even without external drivers. However, we are particularly interested in ``external" turbulence for which the driving sources do not directly depend on dust dynamics, such as the magnetorotational instability (MRI, \citealt{Balbus1991}), or the numerous hydrodynamic instabilities (see \citealt{Lesur2023PPVII} for a recent review).} into account. Observations show that, although PPDs are weakly turbulent compared to predictions for fully-ionized accretion disks, there is evidence for a non-zero level of turbulence that varies substantially between disks. Observations of turbulent line broadening derive an upper limit of turbulent velocity to sound speed ratio $\delta v/c_s \sim 0.01$ for HD 163296 (\citealt{Flaherty2017}; see also \citealt{Flaherty2018} and \citealt{Flaherty2020} for other disks), whereas other disks (DM Tau -- \citealt{Flaherty2020} and IM Lup -- \citealt{Paneque-Carreno2023}) show $\delta v/c_s \sim 0.3$ above 1--2$H$ ($H$ being vertical scale height of gas) away from the mid-plane.
Observations sensitive to the vertical settling of dust grains suggest even weaker turbulence close to mid-plane in Class II disks (i.e., $\delta v/c_s \lesssim 0.003$ in the outer regions of Oph163131 disk, assuming a turnover timescale of turbulent eddies similar to the local orbital timescale; \citealt{VillenaveHighSettling}), or more modest turbulence in Class I disks (e.g., $\delta v/c_s \gtrsim 0.01$ in the outer regions of IRAS04302 disk under the same assumption for the turnover timescale and assuming $\tau_s=0.01$; \citealt{VillenaveModestSettling}). Clearly, the strength of turbulence (at least as inferred from these observations) varies between disks. It is therefore crucial that we understand the effect of turbulence on the SI clumping criterion and planetesimal formation, across a range of turbulence levels from the minimum set by the SI itself up to at least the levels inferred in DM~Tau and IM Lup.

Second, in previous studies of the clumping boundary, particle self-gravity has been neglected, with the threshold for planetesimal formation being instead defined by whether or not the maximum particle density would exceed the Hill density (a condition often referred to as ``strong clumping"; \citetalias{Li21}). This choice is largely justified since in the presence of very weak turbulence (such as that produced by the SI itself), the Hill criterion is likely a sufficient condition for collapse; a particle cloud (i.e., local particle overdensity) within a narrow filament needs to be sufficiently dense to be stable against tidal shear. However, if turbulence becomes important, gravity has to overpower not only tidal shear but also the diffusion to collapse particle clouds under their own weight. Thus, as both turbulent diffusion and tidal shear can counteract self-gravity, a Toomre-like criterion for gravitational instability becomes a more appropriate criterion for collapse, with diffusion playing a similar role to pressure \citep{Gerbig20, Klahr20, GerbigLi2023}. 

These considerations strongly motivate an inclusion of both external turbulence {\it and} particle self-gravity in determining the clumping boundary. Thus, in this paper, we incorporate both the self-gravity of particles and forced turbulence into 3D gas-plus-particle numerical simulations. We carry out 41 simulations in order to quantify the clumping boundary, here defined as the boundary above which gravitationally bound clumps form in our simulations.

We note that a number of previous works have addressed particle concentration and planetesimal formation under the influence of turbulence. For example, \citet{Chen20} and \citet{Orkan20} analytically show that even moderate level of turbulence (i.e., $\alpha_{\textrm{SS}}$ $\gtrsim 10^{-4}$; where $\alpha_{\textrm{SS}}$ is standard Shakura-Sunyaev turbulent viscosity parameter; \citealt{SS73}) can suppress linear growth of the SI. The notion that turbulence may hinder the SI was numerically confirmed by nonlinear SI simulations with externally driven hydrodynamic turbulence \citep{Gole20}. Moreover, there have been number of studies that investigate the effect of turbulence driven by (magneto-)hydrodynamical instabilities present in PPDs, such as the MRI \citep{Johansen2007Rapidformation, Yang2018, Xu22DustSettling} or the vertical shear instability (VSI; \citealt{Nelson2013}; see \citealt{SchaferVSIandSI} for the VSI and SI study), on the particle concentration. The main takeaway from these works is particle concentration can occur in large-scale features (such as localized concentrations of gas pressure) that emerge naturally from the (magneto-)hydrodynamic proccessat at work. We will return to a discussion of these results later in this manuscript, but for now it is worth mentioning that these numerical studies explored a relatively narrow region in parameter space in terms of $\tau_s$ and $Z$. Thus, the influence of turbulence has yet to be unveiled in a broader parameter space. Exploring such a parameter space with turbulence (and particle self-gravity) is the goal of this paper.

This paper is organized as follows. We describe our numerical approach in Section \ref{sec:method}. In Section \ref{sec:results:effect_of_turb}, we examine the effect of turbulence on particle concentration. Section \ref{sec:results:pltfrmtn} is dedicated to the simulations with the particle self-gravity included. Thus, it is in this section where we present a new clumping boundary. We demonstrate the effect of particle feedback in Section \ref{sec:results:feedbackhp} and its influence on vertical profiles of particle density in Section \ref{sec:results:verticalprofile}. Finally, we discuss and summarize our results in Sections \ref{sec:Discussion} and \ref{sec:summary}, respectively.


\section{Method}\label{sec:method}
\subsection{Numerical Method}\label{sec:method:numerical_method}
We use the {\sc Athena}  hydrodynamics+particle code \citep{Stone08,Stone2010, BaiStone10} to perform 3D simulations of the SI. We artificially force turbulence (see Section \ref{sec:method:forcing} for details) in a vertically stratified shearing box, which is a co-rotating patch of a disk sufficiently small such that the domain can be treated in Cartesian coordinates without curvature effects (see e.g., \citealt{Hawley95}).  More specifically, we use a local reference frame at a fiducial radius $R_0$ that rotates at angular frequency $\Omega$. The equations of gas and particles are written in Cartesian coordinates $(\hat{\vecx},\hat{\vecy},\hat{\vecz})$, where $\hat{\vecx},\hat{\vecy}$, and $\hat{\vecz}$ denote unit vectors pointing to radial, azimuthal, and vertical directions, respectively, and the local Cartesian frame is related to the cylindrical coordinate of the disk ($R, \phi, z'$) by $x=R-R_0$, $y=R_0\phi$, and $z = z'$. 

Solid particles are included and treated as Lagrangian super-particles, each of which is a statistical representation of a much larger number of particles with the same physical properties. The particles are coupled to the gas via aerodynamic forces. We adopt periodic boundary conditions in the azimuthal and vertical directions, while shearing periodic boundary conditions \citep{Hawley95} are used in the radial direction.

To solve the hydrodynamic fluid equations, we use the unsplit Corner Transport Upwind (CTU) integrator \citep{Colella1990}, third-order spatial reconstruction \citep{Colella1984}, and an HLLC Riemann solver \citep{Toro06}. The hydrodynamic equations  in the shearing box approximation are 
\begin{equation}\label{eq:eq1}
    \frac{\partial{\rhog}}{\partial{t}} + \nabla \cdot (\rho_g \textit{\textbf{u}}) = 0,
\end{equation}

\begin{equation}\label{eq:eq2}
    \begin{split}
    \frac{\partial({\rhog \vecu})}{\partial{t}} + \nabla \cdot (\rho_g \vecu \vecu + P \vecI) & = \\
    \rhog \left( 3\Omega^2 \vecx - \Omega^2 \vecz  +2 \vecu  \times \vecOmega \right) 
    & + \rhop \frac{(\vecv - \vecu)}{t_{\rm stop}} +  \rhog\fturb,  
    \end{split}
\end{equation}

\begin{equation}\label{eq:eq3}
    P = \rhog c_s^2.
\end{equation}

\noindent
where $\rhog$ is the gas mass density, $\vecu$ is the gas velocity, and $c_s$ is the sound speed.  
In the momentum equation (Equation \ref{eq:eq2}), $\vecI$ is the identity matrix, and $P$ is the gas pressure defined in Equation (\ref{eq:eq3}). On the right hand side of Equation (\ref{eq:eq2}), the first three terms are radial tidal forces (gravity and centrifugal), vertical gravity, and Coriolis force, respectively. The penultimate term denotes the back-reaction from the particles to the gas where $\rhop$, $\vecv$, and $t_{\rm{stop}}$ are the particle mass density, particle velocity, and stopping time of particles, respectively. The last term $\fturb$ is the forcing term for turbulence (see Section \ref{sec:method:forcing}). Note that we use an orbital advection scheme \citep{Masset2000}, in which the Keplerian shear $\vecu_{K} \equiv -(3/2)\Omega x \hat{\vecy}$ is subtracted and we only evolve the deviation $\vecu \equiv \vecu_{\rm tot} - \vecu_{K}$ where $\vecu_{\rm tot}$ is the total gas velocity. The last of the above equations is our isothermal equation of state, which we assume here for simplicity; thus, $c_s$ is constant in space and time.

The equation of motion for particle $i$ (out of $N_{\textrm{par}}$ total particles) is written as 
\begin{equation}\label{eq:eq4}
    \begin{split}
    \frac{d \vecv_i}{dt} & =  3\Omega^2\vecx_i - \Omega^2\vecz_i + 2 \vecv_i \times \vecOmega - \frac{\vecv_i - \vecu}{t_{\rm stop}}  \\ 
    & - 2\eta u_K \Omega \hat{\vecx} + \textbf{a}_g
    \end{split}
\end{equation}
\noindent
and solved with a semi-implicit integrator \citep{BaiStone10}. The orbital advection scheme, which is applied to the gas dynamics, is implemented for the particles as well. In Equation (\ref{eq:eq4}), the first, the second, and the third terms on the right hand side are radial tidal forces (again, gravity and centrifugal), vertical gravity, and the Coriolis force, respectively. The fourth term is the drag force on individual particles. The second-to-last term represents a constant inward acceleration of particles that allow them to drift radially inward in the presence of the radial gas pressure gradient (see \citealt{BaiStone10}). In our treatment (and again following \citealt{BaiStone10}), the parameter $\eta$ is the fraction of the Keplerian velocity by which the orbital velocity of particles is effectively {\it increased} (i.e., made super-Keplerian), while the gas stays at Keplerian velocities; the frame is effectively boosted by an amount $\eta u_K$, which won't change the essential physics. The last term is the particle self-gravity, which is calculated by solving the following Poisson equation:
\begin{equation} \label{eq:Poisson}
    \nabla^2 \Phi = 4\pi G\rho_p,
\end{equation}
where $G$ is the gravitational constant. We use fast Fourier transforms (FFT) to solve the Poisson  equation for the gravitational potential (see \citealt{Simon2016} for more details on the implementation of the Poisson solver) from which we then calculate the self-gravitational acceleration ($\textbf{a}_g = - \nabla \Phi$). Gas self-gravity is ignored in our simulations.

We use the  triangular-shaped cloud (TSC) method outlined in \cite{BaiStone10} to interpolate the gas velocity at the grid cell centers to the particle locations for the calculation of the gas drag term in Equation (\ref{eq:eq4}). The same interpolation scheme is used to map momenta from the particle locations to the grid cell centers to calculate the back-reaction term on the gas in Equation (\ref{eq:eq2}). The TSC method is also used to calculate $\textbf{a}_g$; the mass density of particles is mapped to grid cell centers to solve Equation (\ref{eq:Poisson}) and calculate $\textbf{a}_g$, which is then interpolated back to the location of particles.

\subsection{Initial Conditions and Parameters} \label{sec:method:initial_conditions} 
Due to the typical length scales of the SI being much smaller than the vertical gas scale height ($H = c_s/\Omega$), our domain size is necessarily smaller than shearing box simulations of other processes, such as magnetically driven turbulence (see, e.g., \citealt{Hawley95}).  More specifically, our domain size is  $(L_x,L_y,L_z)$ = $(0.4,0.2,0.8)H$, where $L_x$, $L_y$, and $L_z$ are radial (length), azimuthal (width), and vertical (height) sizes of the boxes, respectively. For the runs with smaller $\tau_s$ and/or stronger forced turbulence, we increase $L_z$ to reduce the number of particles crossing the vertical boundaries (see Table \ref{tab:runlist}). The number of grid cells in each direction is $(N_x,N_y,N_z)$ = $(256,128,512)$, equating to 640 grid cells per $H$ in each direction, respectively.  We adjust the vertical resolution for the taller boxes so as to maintain an equivalent resolution to 640 cells per $H$.

The dynamics of gas and particles in our simulations are determined by five dimensionless parameters. We focus on the four parameters relevant to the particles here and describe the other one characterizing the forced turbulence in the next subsection.

First, the aerodynamic coupling of the particles to the gas is controlled by the dimensionless stopping time
\begin{equation}\label{eq:eqtaus}
    \tau_s \equiv t_{\textrm{stop}}\Omega,
\end{equation}
which also represents particle size ($t_{\textrm{stop}}$ is proportional to the grain size; see, e.g., equation 1.48a in \citealt{2013pss3.book....1Y} for the formula of $t_{\textrm{stop}}$ used in this work.). In each simulation, all particles are the same size, but we vary $\tau_s$ in different simulations, exploring values of $\tau_s$ ranging from 0.01 to 0.1. For reference, these $\tau_s$ values correspond to particle sizes of millimeters to centimeters at 50 AU in a protoplanetary disk with reasonable choices for disk mass, disk size, etc. (e.g., \citealt{Carrera2021}), though there is some variation in these numbers depending on the disk model employed. 

Second, the abundance of the particles relative to the gas is characterized by the ratio of initial surface density of particles ($\Sigma_{p0}$) to that of gas ($\Sigma_{g0} \equiv  \sqrt{2\pi}\rho_{g0}H$):
\begin{equation}\label{eq:eqZ}
    Z \equiv  \frac{\Sigma_{p0}}{\Sigma_{g0}},
\end{equation}
where $\rho_{g0}$ is the initial gas density. The surface density ratio sets the strength of particle feedback on the gas as well (Equation \ref{eq:eq2}). We consider multiple $Z$ values spanning from 0.015 to 0.4.  This particular range of values is motivated by the $Z$ values needed to produce planetesimals for a given degree of turbulence; see below.

Third, a global radial pressure gradient is parameterized by 
\begin{equation}\label{eq:eqPi}
    \Pi \equiv \frac{\eta u_K}{c_s},
\end{equation}
which measures the strength of the headwind and drives radial drift for solids.  In our current work, we fix $\Pi = 0.05$ based on previous work (see, e.g., \citealt{BaiStone2010ApJ,Carrera2015,Sekiya2018} for more information) and to maintain an economical number of simulations. 

Fourth, we control the strength of the particle self-gravity by using the dimensionless parameter
\begin{equation}\label{eq:eqtildeG}
    \tilde{G} \equiv \frac{4\pi G\rho_{g0}}{\Omega^2} = \sqrt{\frac{8}{\pi}}\frac{1}{Q}.
\end{equation}
Here, $Q$ is the Toomre parameter \citep{Toomre1964} for the gas disk. We fix $\widetilde{G} = 0.05$ ($Q \simeq 32$), which allows us to compare our results to previous numerical studies of the SI. The assumed $Q$ value gives the Hill density\footnote{We acknowledge that Equation (\ref{eq:Hill}), which we refer to as `Hill density', is also termed `Roche density' (e.g., \citealt{Yang2017}, \citetalias{Li21}). However, we opt to use `Hill density' throughout this paper since the Hill criterion is more suitable for the stability analysis of a particle cloud that is in orbital motion than Roche criterion, which assumes neither rotation nor orbital motion.  We refer the reader to Appendix B of \citet{Klahr20} for details on the two criteria.} 

\begin{equation}\label{eq:Hill}
    \rho_H = 9\sqrt{\frac{\pi}{8}}Q \simeq 180\rho_{g0}. 
\end{equation}

We set the number of particles as follows: 
\begin{equation}\label{eq:Npar}
    N_{\textrm{par}} = n_p N_{\textrm{cell}} \approx 1.68\times 10^7 \frac{L_xL_yL_z}{(0.4 \times 0.2 \times 0.8)H^3},
\end{equation}

\noindent
where $N_{\textrm{cell}}$ is the total number of cells. The prefactor on the right hand side results from setting $n_p = 1$ as the number of particles per cell. \citet{BaiStone10} show that setting $n_p$ to 1 is necessary to accurately capture density distributions of particles in unstratified SI simulations. However, as \citetalias{Li21} pointed out, the \textit{effective} particle resolution will be $> 1$ for vertically stratified simulations since particle settling leads to a particle layer thickness $H_p < L_z$. Since our simulations include forced turbulence, the effective particle resolution is not known in advance. Nonetheless, we expect that the effective resolution is higher than 1 for all of our simulations since  $H_p < L_z$ is always satisfied in our simulations.  Furthermore, we test how the effective particle resolution impacts our results in Section~\ref{sec:discussion:caveats:numpar}.

The gas is initialized as a Gaussian density profile with vertical thickness $H$ in hydrostatic balance. The particles are initially  distributed with a Gaussian profile about the mid-plane in the vertical direction; the scale height of this Gaussian is $0.02H$. In the horizontal directions, initial positions of particles are randomly chosen from uniform distribution. The initial velocities of the gas are zero (with Keplerian shear subtracted), whereas the particles have initial azimuthal velocity of $\eta u_K = 0.05c_s$. Importantly, the particles are not initially in an equilibrium state. This is because particles have an initial azimuthal velocity causing them to drift radially inward (and we do not assume the Nakagawa–Sekiya–Hayashi equilibrium condition; \citealt{Nakagawa1986}). Moreover, forced turbulence already exists from the initialization of particles (next section), and the turbulent diffusion will not immediately counterbalance the vertical gravity from a host star.  All of our simulations are shown (with associated relevant parameters) in Table \ref{tab:runlist}.

\begingroup 
\setlength{\medmuskip}{0mu} 

\begin{deluxetable*}{ccccccccccccc}
\tablecolumns{13}
\tabletypesize{\scriptsize}
\tablecaption{List of simulations and time-averaged quantities \label{tab:runlist}}
\tablewidth{0.5\textwidth}
\tablehead{
\colhead{Run} & 
\colhead{$\tau_s$} & 
\colhead{$Z$} &     
\colhead{$\alphaD$} &
\colhead{$L_x \times L_y \times L_z$} &
\colhead{$N_x \times N_y \times N_z$}  &
\colhead{$N_{\textrm{par}}$} & 
\colhead{$t_{\rm{sg}}$} & 
\colhead{Collapse?}  & 
\colhead{$\sigmapz$}  &
\colhead{$\rho_p(z=0)$} &
\colhead{$\rho_{p,\rm{max}}$} &
\colhead{[$t_s,t_e$]}  \\
\colhead{} &
\colhead{} & 
\colhead{} & 
\colhead{} & 
\colhead{$H^3$} &
\colhead{} &
\colhead{} &
\colhead{$\Omega^{-1}$} &
\colhead{} & 
\colhead{$H$} &
\colhead{$\rho_{g0}$} &
\colhead{$\rho_{g0}$} &
\colhead{$\Omega^{-1}$} \\
\colhead{$(1)$} &
\colhead{$(2)$} &
\colhead{$(3)$} &
\colhead{$(4)$} &
\colhead{$(5)$} &
\colhead{$(6)$} &
\colhead{$(7)$} &
\colhead{$(8)$} &
\colhead{$(9)$} &
\colhead{$(10)$} &
\colhead{$(11)$} &
\colhead{$(12)$} &
\colhead{$(13)$}
}
\startdata
T1Z4A4  & 0.01 & 0.04 & $10^{-4}$ & $0.4 \times 0.2 \times 1.2$ & $256 \times 128 \times 768$ & $\approx 2.52 \times 10^{7}$ & 400 & N & 0.047 & 1.480 & 3.066 & [200,400]  \\
T1Z6.5A4  & 0.01 & 0.065 & $10^{-4}$ & $0.4 \times 0.2 \times 1.2$ & $256 \times 128 \times 768$ & $\approx 2.52 \times 10^{7}$ & 400 & Y & 0.037 & 4.958 & 50.642 & [200,328] \\
T1Z8A4 & 0.01 & 0.08  & $10^{-4}$ & $0.4 \times 0.2 \times 1.2$ & $256 \times 128 \times 768$  & $\approx 2.52 \times 10^{7}$ & 400 & Y & $\cdots$ & $\cdots$ & $\cdots$ & $\cdots$ \\
T1Z10A4 & 0.01 & 0.1  & $10^{-4}$ & $0.4 \times 0.2 \times 1.2$ & $256 \times 128 \times 768$  & $\approx 2.52 \times 10^{7}$ & 400 & Y & $\cdots$ & $\cdots$ & $\cdots$ & $\cdots$ \\
T1Z10A3.5 & 0.01 & 0.1  & $10^{-3.5}$ & $0.4 \times 0.2 \times 1.2$ & $256 \times 128 \times 768$  & $\approx 2.52 \times 10^{7}$ & 300 & N  & 0.059 & 3.493 & 12.050 & [220,300] \\
T1Z10A3 & 0.01 & 0.1  & $10^{-3}$ & $0.4 \times 0.2 \times 2.0$ & $256 \times 128 \times 1280$  & $\approx 4.19 \times 10^{7}$ & $\cdots$ & N* & 0.147 & 0.739 & 2.180 & [150,356] \\
T1Z12.5A3.5 & 0.01 & 0.125  & $10^{-3.5}$ & $0.4 \times 0.2 \times 1.2$ & $256 \times 128 \times 768$  & $\approx 2.52 \times 10^{7}$ & 400 & Y & 0.052 & 7.673 & 47.378 & [200,400] \\
T1Z15A3.5 & 0.01 & 0.15  & $10^{-3.5}$ & $0.4 \times 0.2 \times 1.2$ & $256 \times 128 \times 768$  & $\approx 2.52 \times 10^{7}$ & 400 & Y & $\cdots$ & $\cdots$ & $\cdots$ & $\cdots$ \\
T1Z20A3.5 & 0.01 & 0.2  & $10^{-3.5}$ & $0.4 \times 0.2 \times 1.2$ & $256 \times 128 \times 768$  & $\approx 2.52 \times 10^{7}$ & 400 & Y & $\cdots$ & $\cdots$ & $\cdots$  & $\cdots$ \\
T1Z20A3 & 0.01 & 0.2  & $10^{-3}$ & $0.4 \times 0.2 \times 2.0$ & $256 \times 128 \times 1280$  & $\approx 4.19 \times 10^{7}$ & 250 & N & 0.106 & 2.577 & 5.725  & [100,250] \\
T1Z25A3 & 0.01 & 0.25  & $10^{-3}$ & $0.4 \times 0.2 \times 2.0$ & $256 \times 128 \times 1280$  & $\approx 4.19 \times 10^{7}$ & 200 & Y & 0.094 & 4.483 & 10.443 & [100,200] \\
T1Z30A3 & 0.01 & 0.3  & $10^{-3}$ & $0.4 \times 0.2 \times 2.0$ & $256 \times 128 \times 1280$  & $\approx 4.19 \times 10^{7}$ & 400 & Y & $\cdots$ & $\cdots$ & $\cdots$ & $\cdots$ \\
T1Z40A3 & 0.01 & 0.4  & $10^{-3}$ & $0.4 \times 0.2 \times 2.0$ & $256 \times 128 \times 1280$  & $\approx 4.19 \times 10^{7}$ & 400 & Y & $\cdots$ & $\cdots$ & $\cdots$ & $\cdots$\\
\hline
T1.3Z8A4 & 0.013 & 0.08  & $10^{-4}$ & $0.4 \times 0.2 \times 1.2$ & $256 \times 128 \times 768$  & $\approx 2.52 \times 10^{7}$ & 400 & Y & $\cdots$ & $\cdots$ & $\cdots$ & $\cdots$\\
\hline
T2Z4A4 & 0.02 & 0.04  & $10^{-4}$ & $0.4 \times 0.2 \times 0.8$ & $256 \times 128 \times 512$ & $\approx 1.68 \times 10^{7}$ & 400 & N & 0.029 & 2.473 & 9.870 & [100,400] \\
T2Z5A4 & 0.02 & 0.05  & $10^{-4}$ & $0.4 \times 0.2 \times 0.8$ & $256 \times 128 \times 512$ & $\approx 1.68 \times 10^{7}$ & 400 & N & 0.025 & 4.073 & 39.036 & [100,370] \\
T2Z8A4 & 0.02 & 0.08  & $10^{-4}$ & $0.4 \times 0.2 \times 0.8$ & $256 \times 128 \times 512$ & $\approx 1.68 \times 10^{7}$ & 400 & Y & $\cdots$ & $\cdots$& $\cdots$ & $\cdots$ \\
\hline
T3Z2A4 & 0.03 & 0.02  & $10^{-4}$ & $0.4 \times 0.2 \times 0.8$ & $256 \times 128 \times 512$ & $\approx 1.68 \times 10^{7}$ & $\cdots$ & N* & 0.028 & 0.803 & 2.376 & [250,600] \\
T3Z3A4 & 0.03 & 0.03  & $10^{-4}$ & $0.4 \times 0.2 \times 0.8$ & $256 \times 128 \times 512$ & $\approx 1.68 \times 10^{7}$ & 300 & N & 0.024 & 1.737 & 7.639 & [100,300] \\
T3Z4A4 & 0.03 & 0.04  & $10^{-4}$ & $0.4 \times 0.2 \times 0.8$ & $256 \times 128 \times 512$ & $\approx 1.68 \times 10^{7}$ & 300 & Y & 0.021 & 3.076 & 25.405 & [100,300] \\
T3Z45A3.5 & 0.03 & 0.045  & $10^{-3.5}$ & $0.4 \times 0.2 \times 0.8$ & $256 \times 128 \times 512$ & $\approx 1.68 \times 10^{7}$ & 400 & N& 0.041 & 1.447 & 7.942 & [100,400] 
 \\
T3Z5A4 & 0.03 & 0.05  & $10^{-4}$ & $0.4 \times 0.2 \times 0.8$ & $256 \times 128 \times 512$ & $\approx 1.68 \times 10^{7}$& 400 & Y & $\cdots$ & $\cdots$ & $\cdots$ & $\cdots$ \\
T3Z5A3.5 & 0.03 & 0.05  & $10^{-3.5}$ & $0.4 \times 0.2 \times 0.8$ & $256 \times 128 \times 512$ & $\approx 1.68 \times 10^{7}$ & 500 & Y & 0.038 & 1.848 & 8.412 & [100,500] \\
T3Z6A3.5 & 0.03 & 0.06  & $10^{-3.5}$ & $0.4 \times 0.2 \times 0.8$ & $256 \times 128 \times 512$ & $\approx 1.68 \times 10^{7}$ & 500 & Y & 0.035 & 2.929 & 29.342 & [100,292] \\
T3Z7A3 & 0.03 & 0.07  & $10^{-3}$ & $0.4 \times 0.2 \times 1.2$ & $256 \times 128 \times 768$ & $\approx 2.52 \times 10^{7}$ & 400 & N & 0.077 & 1.018 & 5.818 & [50,400] \\
T3Z8A3 & 0.03 & 0.08  & $10^{-3}$ & $0.4 \times 0.2 \times 1.2$ & $256 \times 128 \times 768$ & $\approx 2.52 \times 10^{7}$ & 400 & N & 0.072 & 1.279 & 6.508 & [50,400]  \\
T3Z9A3 & 0.03 & 0.09  & $10^{-3}$ & $0.4 \times 0.2 \times 1.2$ & $256 \times 128 \times 768$ & $\approx 2.52 \times 10^{7}$ & 400 & N & 0.070 & 1.518 & 7.164 & [50,400] \\
T3Z10A3 & 0.03 & 0.1  & $10^{-3}$ & $0.4 \times 0.2 \times 1.2$ & $256 \times 128 \times 768$ & $\approx 2.52 \times 10^{7}$ & 300 & Y & 0.066 & 1.881 & 8.216 & [50,300] \\
T3Z20A3 & 0.03 & 0.2  & $10^{-3}$ & $0.4 \times 0.2 \times 1.2$ & $256 \times 128 \times 768$ & $\approx 2.52 \times 10^{7}$ & 300 & Y & $\cdots$ & $\cdots$ & $\cdots$ & $\cdots$\\
\hline
T10Z1.5A4  & 0.1 & 0.015 & $10^{-4}$ & $0.4 \times 0.2 \times 0.8$  & $256 \times 128 \times 512$ & $\approx 1.68 \times 10^{7}$ & 350 & N & 0.015 & 0.994 & 8.726 & [50,350] \\
T10Z2  & 0.1 & 0.02 & No forcing & $0.4 \times 0.2 \times 0.8$  & $256 \times 128 \times 512$ & $\approx 1.68 \times 10^{7}$ & 350 & Y & 0.008 & 2.752 & 41.362 & [50,350] \\
T10Z2A4  & 0.1 & 0.02 & $10^{-4}$ & $0.4 \times 0.2 \times 0.8$  & $256 \times 128 \times 512$ & $\approx 1.68 \times 10^{7}$ & 350 & Y & 0.013 & 1.654 & 36.424 & [100,211]\\
T10Z2A4-$n_p$** & 0.1 & 0.02 & $10^{-4}$ & $0.4 \times 0.2 \times 0.8$  & $256 \times 128 \times 512$ & $\approx 4.61 \times 10^{6}$ & $\cdots$ & $\cdots$ & 0.013 & 1.763 & 69.740 & [100,211]\\
T10Z2A3.5 & 0.1 & 0.02 & $10^{-3.5}$ & $0.4 \times 0.2 \times 0.8$ & $256 \times 128 \times 512$ & $\approx 1.68 \times 10^{7}$ & 350 & N & 0.024 & 0.884 & 10.116 & [50,350] \\
T10Z2A3  & 0.1 & 0.02 & $10^{-3}$ & $0.4 \times 0.2 \times 0.8$   & $256 \times 128 \times 512$ & $\approx 1.68 \times 10^{7}$ & $\cdots$ & N* & 0.050 & 0.425 & 8.631  & [50,600] \\
T10Z2.4A3.5 & 0.1 & 0.024 & $10^{-3.5}$ & $0.4 \times 0.2 \times 0.8$ & $256 \times 128 \times 512$ & $\approx 1.68 \times 10^{7}$ & 350 & Y & 0.022 & 1.244 & 19.944 & [50,350] \\
T10Z3.2A3.5 & 0.1 & 0.032 & $10^{-3.5}$ & $0.4 \times 0.2 \times 0.8$ & $256 \times 128 \times 512$ & $\approx 1.68 \times 10^{7}$ & 350 & Y & $\cdots$ & $\cdots$ & $\cdots$ & $\cdots$ \\
T10Z4A3 & 0.1 & 0.04 & $10^{-3}$ & $0.4 \times 0.2 \times 0.8$ & $256 \times 128 \times 512$ & $\approx 1.68 \times 10^{7}$ & 350 & N & 0.041 & 1.106 & 17.976 & [50,350] \\
T10Z4.5A3 & 0.1 & 0.045 & $10^{-3}$ & $0.4 \times 0.2 \times 0.8$ & $256 \times 128 \times 512$ & $\approx 1.68 \times 10^{7}$ & 500 & Y & 0.040 & 1.308 & 20.246 & [50,500] \\
T10Z5.4A3 & 0.1 & 0.054 & $10^{-3}$ & $0.4 \times 0.2 \times 0.8$ & $256 \times 128 \times 512$ & $\approx 1.68 \times 10^{7}$ & 350 & Y & $\cdots$ & $\cdots$ & $\cdots$ & $\cdots$ \\
\hline  
T30Z2A3.5 & 0.3 & 0.02 & $10^{-3.5}$ & $0.4 \times 0.2 \times 0.8$ & $256 \times 128 \times 512$ & $\approx 1.68 \times 10^{7}$ & 300 & Y & $\cdots$ & $\cdots$ & $\cdots$ & $\cdots$\\ 
\hline 
\enddata 

\tablecomments{Columns: $(1)$: run name (the numbers after T and Z are in units of one hundredth, while those after A are the absolute values of power indices; $(2)$: dimensionless stopping time of particles (see Equation \ref{eq:eqtaus}); $(3)$: initial surface density ratio of particle to gas (see Equation \ref{eq:eqZ}); $(4)$: dimensionless diffusion parameter of turbulence (see Equation \ref{eq:eqaout}); $(5)$; dimensions of the simulation domain in unit of gas scale height; $(6)$: the number of grid cells in each direction; $(7)$: number of particles; $(8)$: when particle self-gravity is switched on; $(9)$: whether or not gravitational collapse of particles occurs (Y for Yes and N for No) $(10)$; time-averaged standard deviation of particles' vertical positions (see Equation \ref{eq:sigmapz}); $(11)$: time-averaged particle density at the disk mid-plane; $(12)$: time-averaged maximum particle density; $(13)$: time interval over which quantities in column (10)-(12) are averaged. We do not report the quantities for runs that have relatively short pre-clumping phases.
All runs have the global radial pressure gradient of $\Pi = 0.05$ and the self-gravity parameter of  $\tildeG = 0.05$. Time evolution, vertical density profile, and final 2D snapshots of particle density for each run listed in the table are available at \href{https://github.com/simon-research-group/Jeonghoon_Public}{https://github.com/simon-research-group/Jeonghoon\_Public}.\\
* We decided not to turn on particle self-gravity in these runs. Nevertheless, we conclude that they are not capable of forming planetesimals even if the gravity is turned on because a run with one step higher Z value but identical $\tau_s$ and $\alphaD$ (e.g., T1Z10A3 vs T1Z20A3) does not lead to planetesimal formation with the self-gravity on. \\
** Since this run is to study the effect of the number of particles on the particle concentration (see Section \ref{sec:discussion:caveats}), we do not include particle self-gravity here.
} 
\end{deluxetable*}

\endgroup

\setlength{\medmuskip}{0mu} 
\begin{deluxetable}{ccccccc}
    \tablecaption{Summary of our forcing parameter and relevant quantities from $\tau_s=0.1,~Z=10^{-5}$ simulations \label{table:gasinfo}}
    \tablewidth{0pt}
    \tablehead{
    \colhead{$\Lambda/(\rho_{g0}^{1/2}c_sH^{3/2})$} &\colhead{$\left.\text{\hspace{10pt}}\right.$} &\colhead{$\alphaD$} &\colhead{$\left.\text{\hspace{10pt}}\right.$}  &\colhead{$\alphaDz$} &\colhead{$\left.\text{\hspace{10pt}}\right.$}  &\colhead{$\sqrt{(\delta u^{'})^2/c_s^2}$}
    }
    \startdata
        $1.3\times10^{-4}$ & & $10^{-4}$ & & $4.0 \times 10^{-5}$ & & 0.016 \\
        $2.5\times10^{-4}$ & & $10^{-3.5}$  & & $1.3 \times 10^{-4}$ & & 0.026\\
        $6.7\times10^{-4} $ & & $10^{-3}$  & & $3.6 \times 10^{-4}$ & & 0.051 \\
    \enddata
    \tablecomments{We report $\alphaD$, $\alphaDz$, and $\delta u'$ values that result from turbulence forced with the interpolated $\Lambda$ (Figure \ref{fig:Lambda_alphaD}). For the SI simulations listed in Table \ref{tab:runlist}, we use these $\Lambda$ values as the initial condition for the forcing amplitude for corresponding $\alphaD$.}
\end{deluxetable}


\subsection{Turbulence Forcing}\label{sec:method:forcing}
As mentioned earlier, we inject turbulence into our simulation domain. Unlike \citet{Gole20} in which turbulence is driven in Fourier space and added to real space via an inverse Fourier transform, we drive turbulence in real space with a cadence of $t_{\rm{drive}} = 0.001\Omega^{-1}$. We compared one of our runs to that in \citet{Gole20}, both of which have the same $\tau_s$, $Z$ and comparable kinetic energy of gas (Run T30Z2A3.5), and found no significant differences. We use a vector potential method to force turbulence, which guarantees that the velocity perturbations introduced into the domain are incompressible (i.e., divergence-free):

\begin{equation}\label{eq:fturb}
    \fturb = \Lambda(\nabla \times \vecA)
\end{equation}
\noindent
where $\vecA$ is the vector potential, and
$\Lambda$ is a forcing amplitude that determines the velocity magnitude of the forced turbulence. The amplitude does not change with time for a constant energy input at every $t_{\rm{drive}}$. The velocity perturbations are obtained by taking a curl of $\vecA$ numerically in a way that guarantees that $\fturb$ is a cell-centered quantity. $\vecA$ and thus $\fturb$ are sinusoidal with a phase that varies randomly every time the forcing is done. We elaborate on the equations for $\vecA$ and the way we handle the shearing-periodic boundary conditions in Appendix \ref{sec:appendix}.  We stress that while our perturbations are initially incompressible, there is no guarantee that the injected turbulence maintains a divergenceless (i.e., incompressible) velocity field.  However, we have verified that in the absence of particles, the divergenceless component of the velocity field accounts for $\sim 99\%$ of the total velocity field (see Appendix \ref{sec:appendix} for details). 

We use a parameter $\alphaD$ to quantify the level of the forced turbulence throughout this paper. The $\alphaD$ is by definition a spatial diffusion of turbulent gas in a dimensionless form as follows:
\begin{equation}\label{eq:eqaout}
    \alphaD \equiv \frac{D_g}{c_sH} \simeq \left(\frac{\delta u'}{c_s}\right)^2\tau_{\rm{eddy}},
\end{equation}
where $D_g \simeq (\delta u')^2t_{\rm{eddy}}$ is a dimensional version of $\alphaD$ representing gaseous diffusion due to velocity fluctuations \citep{Fromang2006, Youdin_Lithwick2007}: $\delta u'_i \equiv \sqrt{\langle u{'}^2_i \rangle - \langle u'_i \rangle^2}$ is the velocity fluctuation of $i$th component ($\langle \cdots \rangle$ means spatial average), the total velocity perturbation is $\delta u' \equiv \sqrt{(\delta u'_x)^2+(\delta u'_y)^2+(\delta u'_z)^2}$, and $\tau_{\rm{eddy}} \equiv t_{\rm{eddy}}\Omega$ is the dimensionless eddy turnover time of turbulence. We adjust $\Lambda$ to obtain our target values of $\alphaD$, which are $10^{-4}$, $10^{-3.5}$, and $10^{-3}$. In the following, we explain how we obtain the forcing amplitude ($\Lambda$) for the three $\alphaD$ values.

First of all, we assume that in the limit $\rho_p \ll \rho_g$, the particle scale height $H_p$ can be determined by the balance between the particle settling and vertical diffusion by turbulence \citep{Youdin_Lithwick2007}:
\begin{equation}\label{eq:eqaz}
     \alphaDz = \tau_s\left[\frac{(H_{p}/H)^2}{1-(H_{p}/H)^2}\right],
\end{equation}
where $\alphaDz$ corresponds to the $\alphaD$ for vertical diffusion only. Second, to obtain $H_p/H$ and the resulting $\alphaDz$, we perform four simulations, each of which has a different $\Lambda$ value, the dimension of $(L_x,L_y,L_z)=(0.4,0.2,0.8)H$, and the resolution of $(N_x,N_y,N_z)=(256,128,512)$. We set $\tau_s = 0.1$ and $Z = 10^{-5}$ in these simulations; the very small $Z$ guarantees $\rho_p \ll \rho_g$ and that the Equation (\ref{eq:eqaz}) holds true (self-gravity of particles is deactivated). Once each simulation reaches a saturated state, we average $H_p/H$ over time for $250\Omega^{-1}$ and obtain $\alphaDz$. Third, since $\alphaDz$ accounts for the diffusion in the vertical direction only, we relate $\alphaDz$ to $\alphaD$ by 
\begin{equation}\label{eq:aoutaz}
    \begin{split}
    \alphaD = \frac{(\delta u'_x)^2+(\delta u'_y)^2+(\delta u'_z)^2}{c_s^2}\tau_{\rm{eddy}} \\
          = \frac{(\delta u'_z)^2}{c_s^2}\tau_{\rm{eddy}}(l^2+m^2+1), \\
          = \alphaDz(l^2+m^2+1),
    \end{split}
\end{equation}
\noindent
where $l \equiv \delta u'_x/\delta u'_z$ and $m \equiv \delta u'_y/\delta u'_z$.  In other words,  $l^2+m^2+1 = 3$ means isotropic turbulence\footnote{We note that the forced turbulence is not perfectly isotropic. The anisotropy is mainly caused by $\delta u_y'$ which is systematically lower than the other two components. We believe this is because eddies with large  $\tau_{\rm{eddy}}$ are more easily destroyed by orbital shear which acts along $y$ direction. If this is true, $\delta u'_y$ can be lower than the other two components since larger eddies contribute to the kinetic energy of turbulence more according to Kolmogorov turbulence theory. It is also possible that since $L_y < L_x$, the narrower side limits sizes of eddies along the y direction and in turn lowers $\delta u'_y$. }. We time-average $l$ and $m$ within the same time interval that we use for time-averaging $H_p/H$. The eddy turnover time ($\tau_{\rm{eddy}}$) is assumed to be isotropic so that $\tau_{\rm{eddy}} = \tau_{\rm{eddy},i}; i=x,y,z$. In this manner, we can calculate $\alphaD$ in each of the four simulations without needing to know $\tau_{\rm{eddy}}$ so that we establish the relation between $\Lambda$ and $\alphaD$. Using the relation, we perform linear interpolation between the data points (i.e., $\alphaD$ at each $\Lambda$) to find the appropriate $\Lambda$ that is expected to produce the desired $\alphaD$ values.

\begin{figure}
    \includegraphics[width=\columnwidth,]{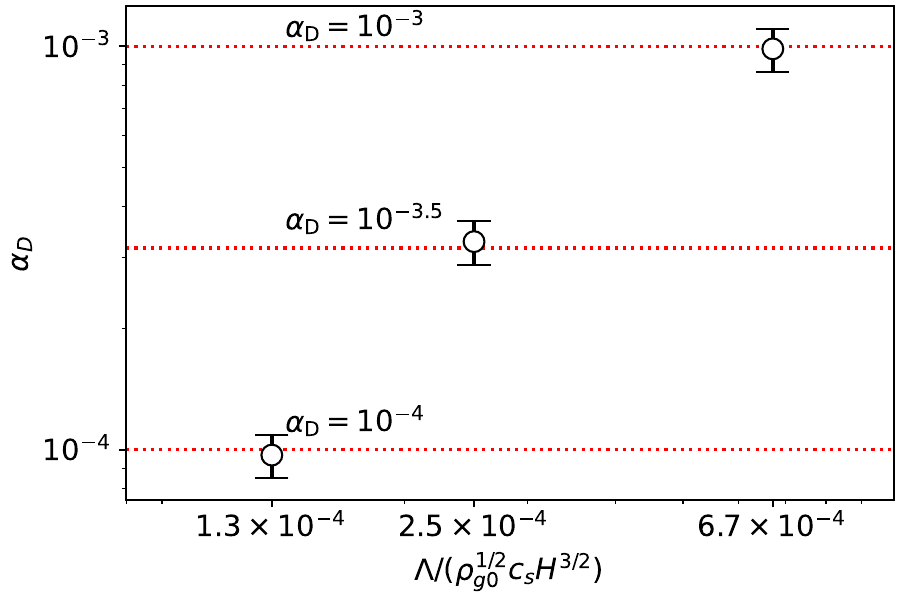}
    \caption{Relationship between $\Lambda$ and $\alphaD$ in simulations with $\tau_s = 0.1, \ Z = 10^{-5}$. The former is an initial condition for a forcing amplitude (Equation \ref{eq:fturb}), while the latter is the parameter (Equation \ref{eq:eqaout}) we use to denote the level of turbulence in our SI simulations. The $\Lambda$ values shown in this plot are obtained by a linear interpolation (see the text for details). Each red horizontal line indicates each $\alphaD$ we choose: $10^{-4}$, $10^{-3.5}$, and $10^{-3}$. The error bar shows $\pm 1$ standard deviation. We confirm that the interpolated $\Lambda$ values result in the desired $\alphaD$ values with high accuracy. 
    }\label{fig:Lambda_alphaD}
\end{figure}

We perform three additional simulations, each of which has $\Lambda$ obtained from the interpolation. Then, we measure $\alphaD$ via the same procedure we describe above to confirm whether the interpolated $\Lambda$ results in the desired values of $\alphaD$. Figure \ref{fig:Lambda_alphaD} shows the resulting $\alphaD$ for each interpolated $\Lambda$ value, which is summarized in Table \ref{table:gasinfo}. The error bar shows $\pm 1$ standard deviation due to the temporal fluctuation of $H_p$ and gas velocities. Clearly, we achieve our desired $\alphaD$ values, denoted by red vertical lines, with very high accuracy. These interpolated $\Lambda$ values are adopted as the initial condition for the forcing amplitude and kept as constant in our SI simulations. However, instead of $\Lambda$, we use $\alphaD$ to denote the level of the forced turbulence to make contact with standard disk dynamics notation. Hence, each of our SI simulations has an unique combination of ($\tau_s$, $Z$, $\alphaD$) as listed in Table \ref{tab:runlist}; ``No forcing" run refers to a simulation where $\fturb = 0$.

We clarify that the $\alphaD$ ($\alphaDz$) measures \textit{bulk (vertical) diffusion} in the gas set by (vertical) velocity fluctuations and the eddy turnover time of the forced turbulence. As noted by previous studies (e.g., \citealt{Youdin_Lithwick2007, Yang2018}), one should not interpret the parameter as $\alpha_{\rm{SS}}$, which is responsible for angular momentum transport due to \textit{turbulent shear stress} in a disk.

Before initializing particles in the SI simulations, we force turbulence with the interpolated $\Lambda$ values up to $t_{\rm{par}}~=~300\Omega^{-1}$ in order to let turbulence fully develop without being affected by particles and also to reach a statistically steady state. At $t=t_{\rm{par}}$, we initialize particles as described in Section \ref{sec:method:initial_conditions}. Before the initialization of particles, we verified that the Fourier spectrum of $(\delta u')^2$ (see Appendix \ref{sec:appendix}) is reasonable with power across a range of scales, and with the most power at the driving scale ($\sim 0.1H$). In the no-forcing run, particles are initialized at the beginning. We use the notation of $t-t_{\rm{par}}$ (in units of $\Omega^{-1}$) for our temporal dimension; particle initialization is done at $t-t_{\rm{par}} = 0$.



\begin{figure*}
    \centering
    \includegraphics[width=\textwidth]{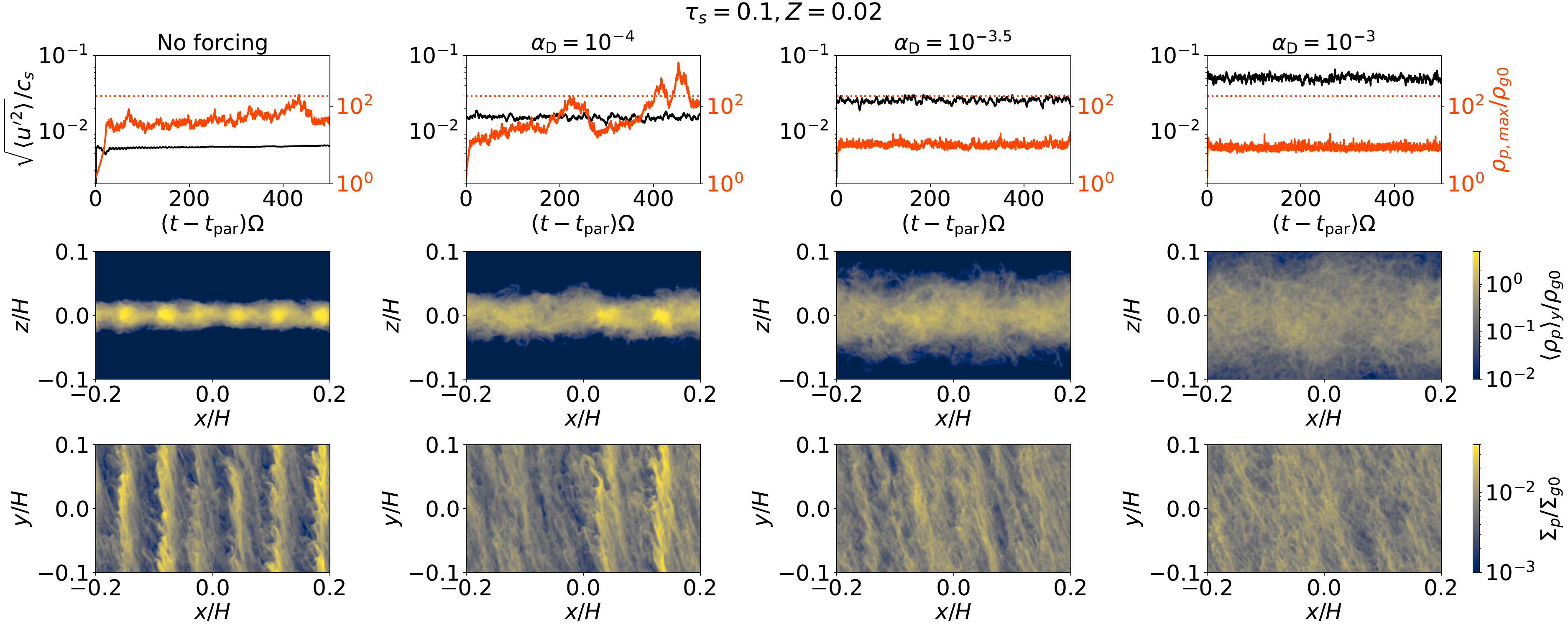}
    \caption{$\it{Top}$: Time evolution of gas rms velocity (black) and maximum density of particles (orange) in simulations with $\tau_s = 0.1, Z = 0.02$. $\it{Middle}$: Azimuthally averaged particle density (side view) zoomed in to the region $z/H \in [-0.1,0.1]$. $\it{Bottom}$: Vertically integrated particle mass density (i.e., surface density of particles; $\Sigma_p$; top view). The snapshots are taken at $t-t_{\rm{par}} = 350\Omega^{-1}$. Note that the middle and the bottom panels have different colorbar scales. From left to right, no-forcing, $\alphaD = 10^{-4}$, $10^{-3.5}$, $10^{-3}$ runs are shown. The horizontal line in each panel on the top marks Hill density (see Equation \ref{eq:Hill}). Particle self-gravity is disabled in the runs presented here. As turbulence becomes stronger from left to right, the particle layer become thicker, and the filaments become more diffused and less well-defined. 
    } 
    \label{fig:tau01-Z002-overview}
\end{figure*}

\section{Results}\label{sec:results}
We present a summary of statistics of our simulations in Table \ref{tab:runlist}. We report temporal averages of particle scale height, maximum particle density, and mid-plane density ratio of the particle and the gas. We adopt a measurement strategy similar to that of \citetalias{Li21} to facilitate comparison with their results: quantities are averaged after the particles have settled from their initial positions and are in a statistical equilibrium (vertically) against turbulent diffusion. The time average is done either before the maximum density first exceeds $(2/3)\rho_{H}$ (termed as pre-clumping phase, \citetalias{Li21}), or before the particle self-gravity is turned on (hereafter, $t_{\rm{sg}}$), whichever occurs first. Table \ref{tab:runlist} presents $t_{\rm{sg}}$ of simulations at which the self-gravity is implemented. We opt not to include the self-gravity in three of our runs as they seem incapable of forming planetesimals. This choice is made when the run with one step higher $Z$ value, but identical $\tau_s$ and $\alphaD$, does not result in planetesimal formation. Since the maximum density of these three runs never reaches $(2/3)\rho_{H}$, the time-averaging is done all the way to the end of the simulations. We exclude from our reported statistics any simulation where strong particle concentrations happens too rapidly. More specifically, we exclude simulations where a quasi-steady state only lasts for a few tens of $\Omega^{-1}$ or is never achieved before the maximum density reaches the threshold.

\subsection{Effect of Turbulence on the Particle Concentration}\label{sec:results:effect_of_turb}
We show the effect of the turbulence on particle dynamics without particle self-gravity before presenting more detailed analysis of our results. Figure \ref{fig:tau01-Z002-overview} shows the results of simulations with $\tau_s = 0.1, Z = 0.02$ but for different $\alphaD$ values. From top to bottom, gas rms velocity (black) and maximum density of particles (orange) as a function of time, azimuthally averaged, and vertically integrated particles densities ($\Sigma_p = \int \rho_p dz$) are shown. The horizontal lines in the top panels denote Hill density, $\rho_{H} \simeq 180\rho_{g0}$.

As can be seen from the top panel, gas rms velocity for $\alphaD = 10^{-4}, 10^{-3.5}$, and $10^{-3}$ levels off at $\sim 0.01c_s$, $\sim 0.03c_s$, and $\sim 0.05c_s$, respectively, while that of the no-forcing run stays only at $\sim 0.006c_s$. These values were calculated via time averaging between $t_s$ and $t_e$, where this time-averaging interval is determined based on the criterion outlined above (also see Table \ref{tab:runlist}). This directly affects the evolution of maximum density of particles; the peak density is very close to or exceeds the Hill density in the no-forcing and $\alphaD = 10^{-4}$ runs, while it saturates at only $\sim 10\rho_{g0}$ in the other two runs due to the stronger turbulence. Interestingly, weak turbulence (i.e., $\alphaD = 10^{-4}$) seems to enhance the particle concentration more than the no-forcing run does. However, given the stochastic nature of clumping, it is hard to draw any firm conclusions about whether $\alphaD = 10^{-4}$ actually produces stronger clumping than the same system without forced turbulence.

The middle and the bottom panels in Figure \ref{fig:tau01-Z002-overview} presents snapshots for each run at $t-t_{\rm{par}} = 350\Omega^{-1}$ that reveal the degrees of particle concentration. First, the side view of the particle layer (i.e., middle panels) clearly shows that the turbulence vertically stirs particles and thus thickens the particle layer. The particle scale heights at $t-t_{\rm{par}} = 350\Omega^{-1}$ are $\sim0.008H$, $\sim0.01H$, $\sim0.02H$, and $\sim0.05H$ for the no-forcing, $\alphaD = 10^{-4}, 10^{-3.5}$, and $10^{-3}$ runs, respectively. Second, as can be seen from the bottom panels, the no-forcing and $\alphaD = 10^{-4}$ runs have azimuthally elongated particle filaments, which implies that the SI is active.  Even though the filaments form in both runs, the latter run has fewer filaments than the former run.  The $\alphaD = 10^{-3.5}$ run shows marginal filament formation, and the $\alphaD = 10^{-3}$ run shows no  evidence for filament formation, instead showing a very diffused particle medium. 

\begin{figure*}
    \centering
    \includegraphics[width=\textwidth]{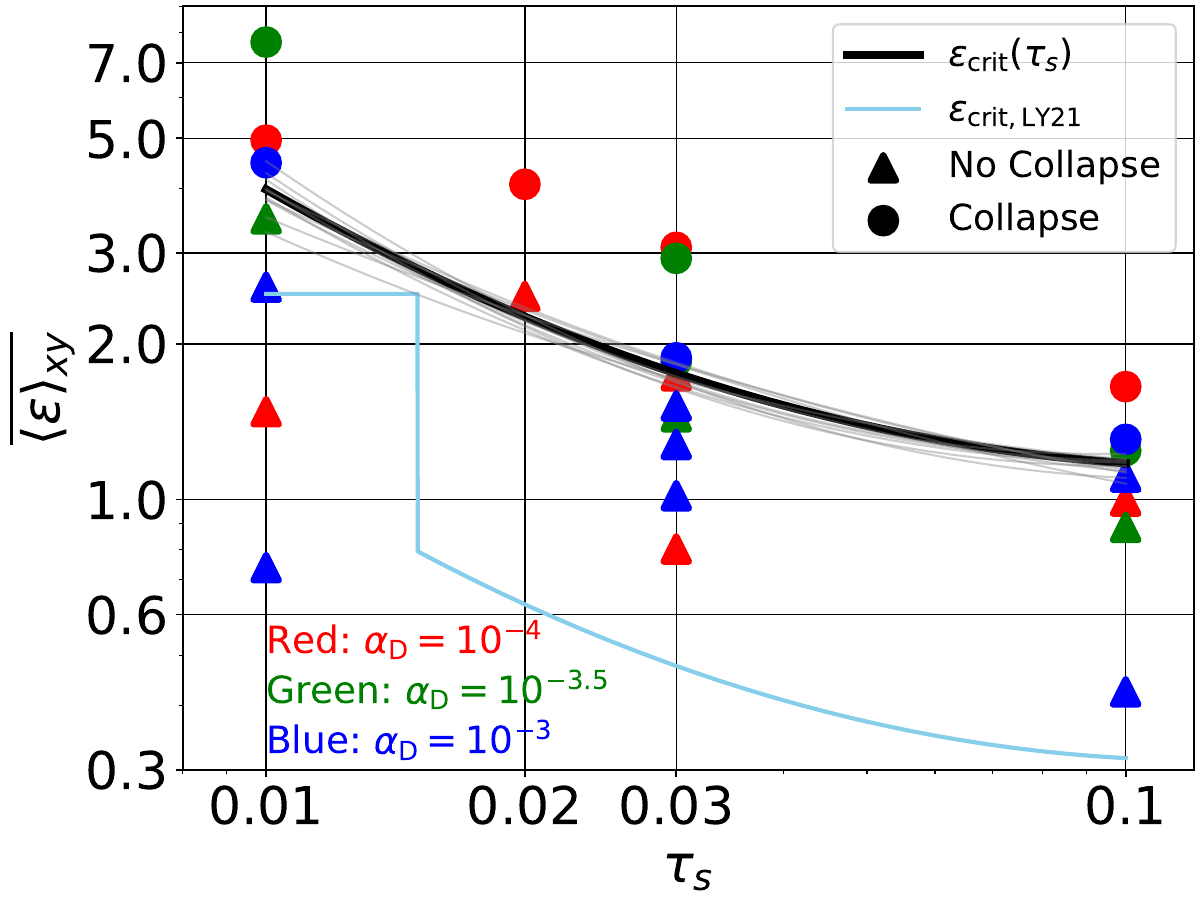}
    \caption{Temporally and horizontally averaged ratio of particle density to gas density at the mid-plane as a function of $\tau_s$ in our simulations. We do not include simulations where the pre-clumping phase is too short to take time-averages. Runs where the collapse occurs are shown as circles, whereas those where the collapse does not occur are shown as triangles. Red, green, and blue represent $\alphaD = 10^{-4}$, $10^{-3.5}$, and $10^{-3}$, respectively, while we did not denote $Z$ values. The black curve denotes the least squares best fit to the critical value of $\epsilon$ as described in Equation (\ref{eq:epscrit}). The thin grey curves are random fits drawn from multivariate normal distribution of $A,B,C$ (Equation \ref{eq:epscrit}) that show uncertainties in the best fit curves. We also include the critical curve from \citetalias{Li21}, depicted as a sky blue curve ($\epscritLY$). The figure shows that $\epscrit$ does not vary with $\alphaD$ (i.e., even changing $\alphaD$, every no-collapse case is below every collapse case), which results in a single critical curve that fits all $\epsilon$ values regardless of $\alphaD$.
    }
    \label{fig:epsmap} 
\end{figure*}

The two-dimensional distributions of particle density in Figure \ref{fig:tau01-Z002-overview} suggest that turbulence can weaken or even completely suppress the SI.  When turbulence is weak or moderate (i.e., $\alphaD = 10^{-4}$ and $10^{-3.5}$), the SI forms elongated filaments. However, the resulting filaments are fewer and less dense than the no-forcing case. Conversely, the $\alphaD = 10^{-3}$ case shows no filamentary structures at all (see the bottom right panel). This could be attributed to very strong vertical stirring that prevents particle settling. It is also possible that the SI is active but that filament formation is overpowered by destructive diffusion. The latter scenario is evident in Runs T1Z10A3.5 and T2Z4A4. Despite these runs showing $\rho_p \gtrsim \rho_g$ at the mid-plane, as indicated in Table \ref{tab:runlist} and Figure \ref{fig:epsmap}, the presence of the turbulent diffusion prevents the SI from forming dense particle filaments, consequently inhibiting planetesimal formation.

\begin{figure*}
    \centering
    \includegraphics[width=\textwidth]{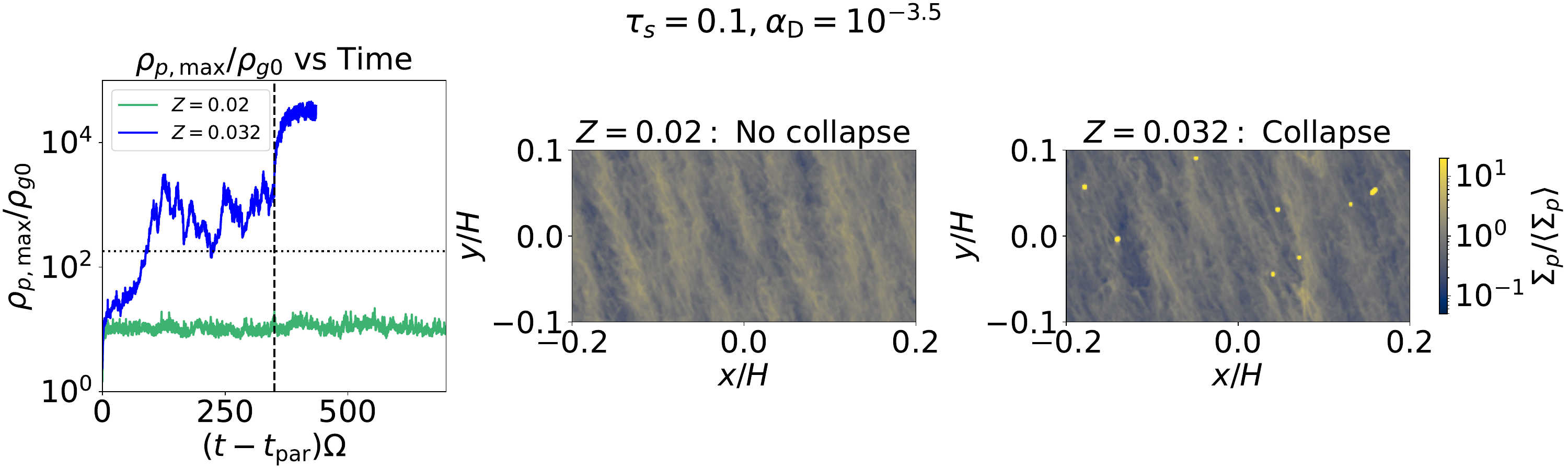}
    \caption{Proof of concept for categorizing runs as ``Collapse" or ``No Collapse". $\it{Left}$: Time evolution of maximum particle density for T10Z2A35 (green) and T10Z3.2A35 (blue) runs. The horizontal and vertical lines indicate Hill density and $t_{\rm{sg}}$, respectively.  $\it{Middle}$: Final snapshot of $\Sigma_p$ divided by its spatial average for T10Z2A3.5 run. $\it{Right}$: Same as the middle panel but for T10Z3.2A3.5 run. We define a run as ``Collapse" if the maximum density drastically increases as shown by the blue curve in the left panel {\it and} if the 2D snapshot shows bound objects (i.e., strong overdensities on very small scales as shown in the right panel).  
    }
    \label{fig:tau01-1e-35-dmax} 
\end{figure*}

\begin{figure*}
    \centering
    \includegraphics[width=\textwidth]{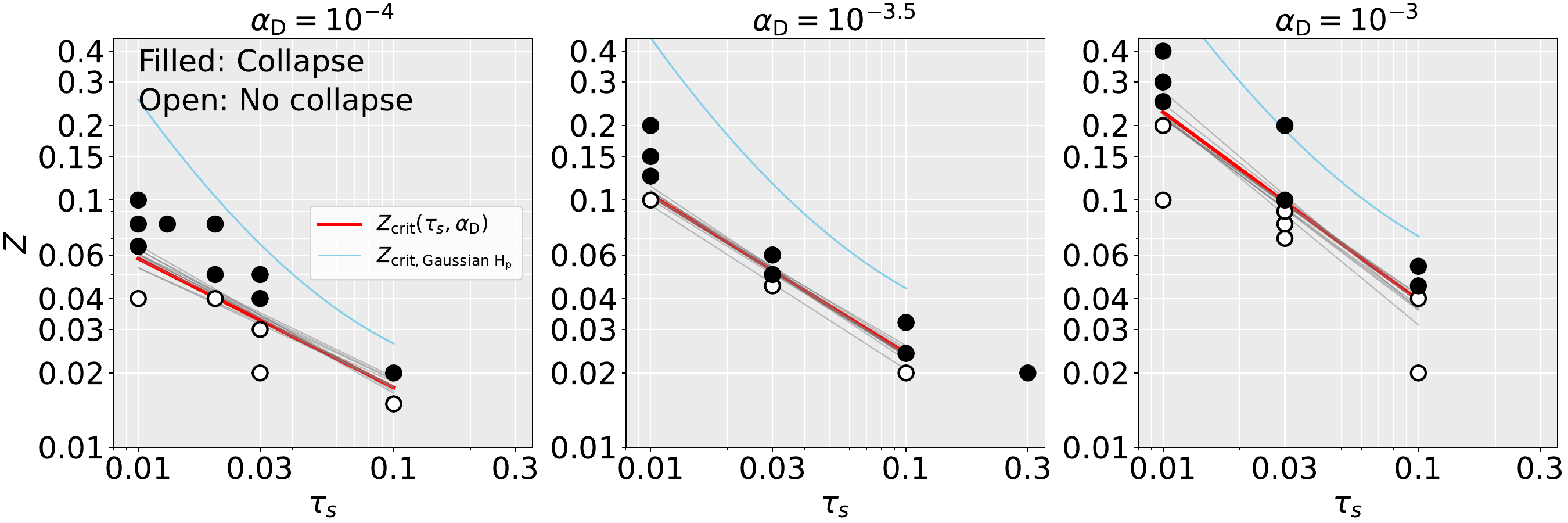}
    \caption{Overview of gravitational collapse in the SI simulations listed in Table \ref{tab:runlist} except Run T10Z2 in which no forcing is applied. Runs where the collapse occurs are shown as filled circles, whereas those where the collapse does not occur are shown as open circles.  From left to right, $\alphaD = 10^{-4}$, $10^{-3.5}$, and $10^{-3}$, respectively. In each panel, we plot two different $\Zcrit$, skyblue being Equation (\ref{eq:LY21Zcrit}) and red being Equation (\ref{eq:ourZcrit}). Both of the curves are fits to the critical Z values but by using different approaches; the red curve is the best fit by multivariate least squares (Equation \ref{eq:ourZcrit} shows the formula), whereas the skyblue one uses the $\epsilon_{\rm crit}$ from Equation~(\ref{eq:epscrit}) to calculate the critical $Z$ assuming that the vertical profile for particle density is Gaussian. The thin grey curves in each panel are random fits drawn from a multivariate normal distribution of $A',B',C',D'$ (Equation \ref{eq:ourZcrit}) values based on their uncertainties (see main text); thus, these curves represent an uncertainty on the $\Zcrit$ curves. Every run shown here has $\Pi = 0.05$ and $\widetilde{G} = 0.05$ (or $Q \sim 32$). We emphasize that $Z_{\rm{crit}}(\tau_s,\alphaD)$ must be used in the range of $\tau_s$ and $\alphaD$ that is given below Equation (\ref{eq:ourZcrit}). Given that the critical Z value for $\tau_s = 0.01$ is $\sim$ 0.02 without external turbulence (e.g., \citetalias{Li21}), external turbulence significantly increases the critical value; $\Zcrit(\tau_s = 0.01,\alphaD) \sim 0.06$, $\sim 0.1$, $\sim 0.2$ for $\alphaD = 10^{-4}$, $10^{-3.5}$, and $10^{-3}$, respectively. A 3D, interactive version of $\Zcrit(\tau_s, \alphaD)$ is available at \href{https://jhlim.weebly.com/research.html}{https://jhlim.weebly.com/research.html}. 
    \label{fig:tauZmap} 
    }
\end{figure*}

\subsection{Planetesimal Formation}\label{sec:results:pltfrmtn}

In this section, we present results from the simulations that incorporate both forced turbulence and particle self-gravity. However, before discussing the results, we describe how we differentiate between ``Collapse" and ``No collapse" runs. The left panel of Figure \ref{fig:tau01-1e-35-dmax} (which corresponds to the run with $\tau_s = 0.1, \ \alphaD = 10^{-3.5}$) shows the maximum particle density as a function of time for $Z = 0.02$ (green) and $Z = 0.032$ (blue) runs, the former being ``No collapse" and the latter being ``Collapse". The horizontal line corresponds to the Hill density, and the vertical line indicates $t_{\rm{sg}}$ in both simulations. The ``Collapse" run shows the maximum density sharply increasing by more than a factor of 10 right after $t_{\rm{sg}}$, which is evidence for the gravitational collapse of particles. On the other hand, the ``No collapse" run shows the steady evolution of the maximum density even though the density slightly increases upon turning on the self-gravity. Furthermore, we investigate the spatial distribution of the particle density as well. The middle and the right panels of the figure show the final snapshots of the surface density of the ``No collapse" and ``Collapse" runs, respectively. The right panel reveals gravitationally bound objects, which we call planetesimals\footnote{As is standard in these types of simulations \cite[e.g.,][]{Simon2016}, our gravity solver prevents collapse of these bound objects below the grid scale.  Thus, it would be more accurate to refer to these objects as diffuse pebble clouds (since we do not have sink particles). However, for simplicity and to make contact with the literature, we use the term planetesimals.}, while the middle panel has no such objects but shows weak filaments have formed. In summary, we consider both the temporal evolution and the spatial distribution of particle density to categorize runs into ``Collapse" and ``No collapse".

\subsubsection{Threshold for Planetesimal Formation: Particle Density at Midplane}\label{sec:results:pltfrmtn:epsilon}
In unstratified SI, the mid-plane density ratio of particle and gas, $\epsilon \equiv \rho_{p}/\rho_{g}(z=0)$, is a crucial parameter. More specifically, when $\tau_s \ll 1$, the linear growth rate of the SI increases with $\epsilon$, with a sharp increase as $\epsilon$ approaches and surpasses unity \citep{YG05}.  Therefore, $\epsilon \gtrsim 1$ is often assumed as a condition for the SI to produce dense clumps of particles. While this could be true in the linear regime of unstratified SI, this is not necessarily true in the fully nonlinear $\it{stratified}$ SI, as pointed out by \citet{Yang2018}. Numerical simulations have shown that critical $\epsilon$ values can deviate from unity depending on $\tau_s$ and other factors. More specifically, \citetalias{Li21} reported a critical $\epsilon$ (for strong clumping to occur) from $\approx 0.3$ to $\approx 3$ depending on the value of $\tau_s$. This is consistent with previous work by \citet{Gole20} in which the critical $\epsilon$ (for planetesimal formation to occur) is $\approx 0.5$ in the presence of external turbulence.  Here, we follow up on this work and further examine the critical $\epsilon$ {\it for planetesimal formation to occur} (since we include particle self-gravity) but for a wider range of parameters than \cite{Gole20}.

Figure \ref{fig:epsmap} shows temporally and horizontally averaged values of $\epsilon$ from the simulations for which we could take time sufficiently long time averages during  the pre-clumping phase (i.e., not every simulation in Table \ref{tab:runlist} is shown). We calculate $\epsilon$ by taking particle and gas densities at $\pm 1$ one grid cell above and below the mid-plane. ``Collapse" and ``No collapse" runs are denoted by circles and triangles, respectively. Red, green, and blue colors denote $\alphaD=10^{-4}$, $10^{-3.5}$, and $10^{-3}$, respectively; we did not color-code $Z$ values. The black curve is the best fit by least squares (see below) to the data marking the approximate location of the critical $\epsilon$ value, above which collapse occurs. In choosing this fit, we assume a quadratic function in log-log space as in \citetalias{Li21} but with different coefficients. More precisely, we find a critical $\epsilon$ of:
\begin{equation}\label{eq:epscrit}
    \log{\epsilon_{\rm{crit}}} \simeq A(\log{\tau_s})^2 + B\log{\tau_s} + C,
\end{equation}
where $A = 0.41, \ B = 0.71, \ C = 0.37$ for all $\alphaD$ values. We also include a fit to the critical $\epsilon$ in \citetalias{Li21} shown as the skyblue curve $(\epscritLY)$.

We calculate this least squares fit as follows. We assume that at a given $\tau_s$ and $\alphaD$, a critical $\epsilon$ falls in the middle between the adjacent no-collapse data point (triangle) and the collapse data point (circle). Moreover, we take the range between the adjacent no-collapse and collapse points as the 95\% confidence interval for the location of $\epscrit$. Therefore, we have a data point (i.e., a critical $\epsilon$) and its associated error at each $\tau_s$ and $\alphaD$. These are then input into a weighted least squares algorithm that accounts for the varying error sizes (heteroscedastic errors). The thin, grey curves in Figure \ref{fig:epsmap} are ten random sample fits, each of whose coefficients are drawn from a multivariate normal distribution of $(A,B,C)$. The width of the distribution in each ``dimension" (i.e., variable) is taken from the covariance matrix from the fit. Those random samples thus provide a sense of the uncertainty in the best fit curve.

Figure \ref{fig:epsmap} has several implications. First, as $\alphaD$ changes, every collapse run remains above every no-collapse run. Thus, the $\epsilon_{\rm{crit}}$ curve precisely cuts between the no-collapse and collapse points  {\it  regardless of $\alphaD$ values}. This implies that $\epsilon_{\rm{crit}}$ varies with $\tau_s$ but not significantly with $\alphaD$. This could potentially be attributed to the use of higher $Z$ values for larger $\alphaD$ at a given $\tau_s$. For instance, at $\tau_s = 0.1$, the $Z$ values of the runs just above the curve are 0.02, 0.024, and 0.045 for $\alphaD = 10^{-4}$, $10^{-3.5}$, and $10^{-3}$, respectively. In other words, although larger $\alphaD$ makes a particle layer thicker, using higher $Z$ adds more mass of particles within the layer, resulting in similar particle densities around the mid-plane (i.e., similar $\epsilon$) in the three $\alphaD$ cases. However, we again emphasize that the accuracy of our fit is limited due to the sparsity of data points across the $\tau_s$ range considered in this study. Consequently, the fit could potentially exhibit variation with $\alphaD$ if additional numerical simulations were to be conducted to further populate parameter space.  

Second, based on Equation (\ref{eq:epscrit}), the values of $\epsilon_{\rm{crit}}$ approximate to 3.98, 1.75, and 1.18 for $\tau_s = 0.01$, 0.03, and 0.1, respectively. These values are several times larger than those from $\epscritLY$\footnote{\citetalias{Li21} found similar $\epscrit$ with \citet{Gole20} at $\tau_s$ = 0.3, whereas our results are inconsistent with \citetalias{Li21}. While this may imply that the two different forcing methods produce inconsistent results, we note that \citet{Gole20} used $\epscrit \approx Z/(H_p/H) \approx Z(\tau_s/\alpha_{\rm{crit}})^{1/2} \sim 0.5$ with $\alpha_{\rm{crit}} = 10^{-3.25}$. This calculation for $\epscrit$ assumes $\rho_p \ll \rho_g$, which is not always the case in our simulations, especially close to the mid-plane where $\epsilon \gtrsim 1$; see Figure \ref{fig:epsmap}}. Moreover, since we have only one simulation at $\tau_s=0.3$, $\epscrit$ at this $\tau_s$ value remains unknown in our work. Moreover, the majority of the data points, regardless of whether or not a corresponding run shows planetesimal formation, are well above $\epscritLY$. A potential explanation for this is that even when particles are able to settle to the mid-plane and form a layer to have $\epsilon \gtrsim 1$, further concentration may be required for them to withstand the turbulence that disperses particles (through aerodynamic coupling with the gas) in all directions. 

Lastly, since $\epsilon_{\rm{crit}}$ we report in this paper is the critical value for \textit{gravitational collapse} of particle clumps rather than just for the SI-induced concentration, the condition that $\epsilon$ simply be greater than unity does not necessarily guarantee gravitational collapse (see \citealt{Gerbig20,Klahr20,Klahr21,GerbigLi2023} for details of the collapse criterion). This is because conditions for gravitational collapse of a particle cloud should be dependent on internal (to the pebble cloud) turbulent diffusion of particles within the cloud as well as its density.

\subsubsection{Threshold for Planetesimal Formation: Critical $Z$}\label{sec:results:pltfrmtn:Zcrit}
Our main results are shown in Figure \ref{fig:tauZmap}. It demonstrates for which parameters ($\tau_s, Z, \alphaD$) planetesimals form. In the figure, filled and open circles correspond to ``Collapse" and ``No collapse" runs, respectively. All simulations in the figure maintain a constant $\Pi$ and $\tildeG$, both of which are 0.05. We do not show Run T10Z2 in the figure, which includes self-gravity but not forced turbulence (see Table \ref{tab:runlist} for the details of this simulation).

The figure demonstrates that planetesimal formation via the SI may be very difficult in the presence of external turbulence. Taking the $\tau_s = 0.01$ cases for example, the critical $Z$ values are $\gtrsim$ 0.06, $\gtrsim$ 0.1, and $\gtrsim$ 0.2 for $\alphaD = 10^{-4}$, $10^{-3.5}$, and $10^{-3}$, respectively. On the other hand, $Z \sim 0.02$ is enough for the SI to produce dense clumps in the absence of external turbulence for this value of $\tau_s$ (\citealt{Carrera2015, Yang2017}, \citetalias{Li21}), which would lead to planetesimal formation if self-gravity of particles was activated (none of these studies included particle self-gravity).

\begin{figure*}
    \centering
    \includegraphics[width=\textwidth]{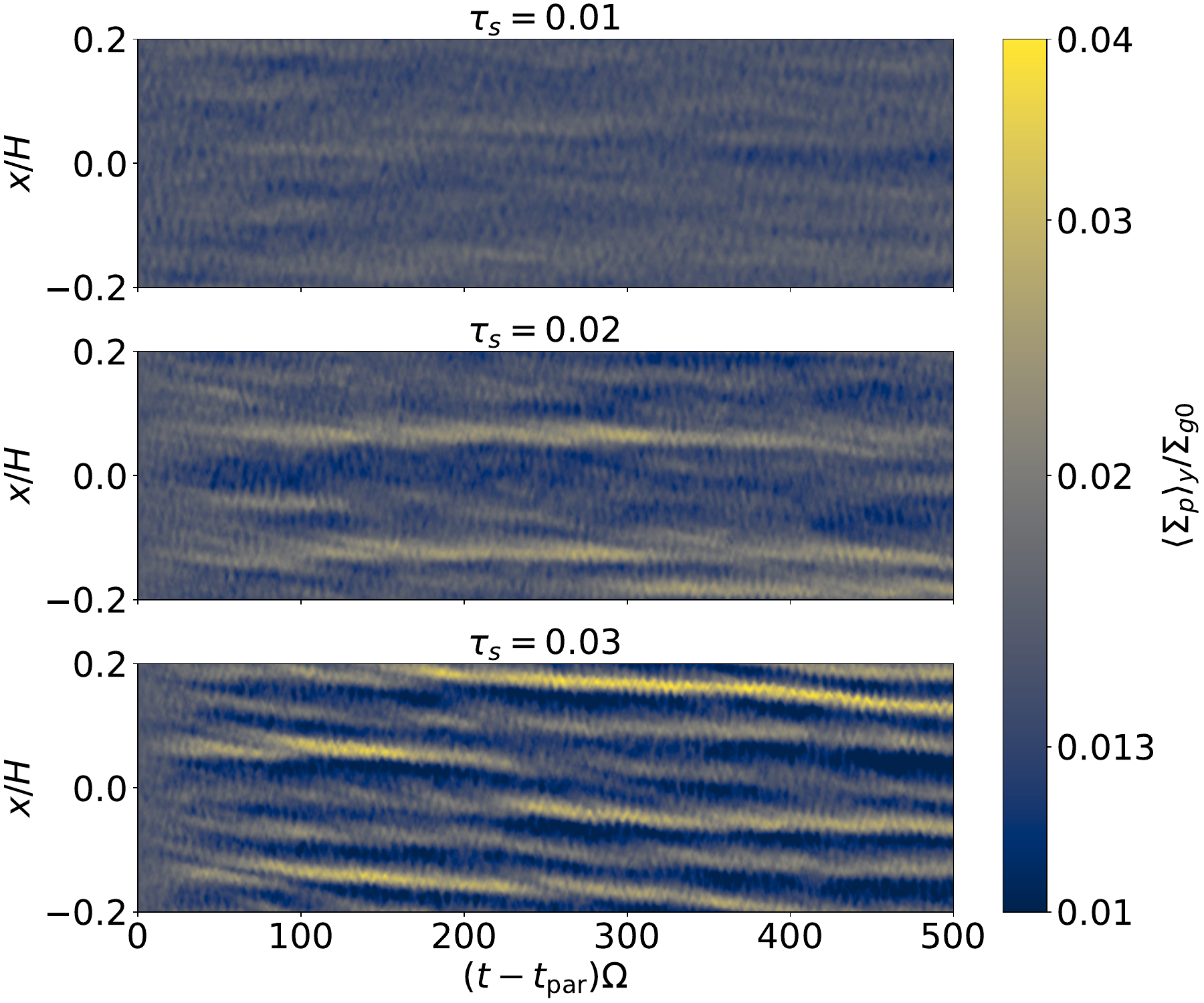}
      \caption{Space-time plots of the particle density that is integrated over $z$ $(\Sigma_p)$ and averaged in $y$ ($\langle \Sigma_p \rangle_y)$. We plot the quantity vs. $x$ and time. From top to bottom, $\tau_s = 0.01$, 0.02, and 0.03, respectively. The runs have the same $Z$ and $\alphaD$, which are 0.04 and $10^{-4}$, respectively. Particle self-gravity is turned off during the time span considered here. The filaments are frequently disrupted unless they are sufficiently dense, resulting in stochastic evolution for all three $\tau_s$ values.
    }
    \label{fig:Zdiscont} 
\end{figure*}

We also plot the critical $Z$ curve {\it assuming a Gaussian scale height} for the particles ($\Zcrithp$, \citetalias{Li21}) and using $\epsilon_{\rm crit}$ as fit by Equation~(\ref{eq:epscrit}) in sky-blue. The curve is given by 
\begin{equation}\label{eq:LY21Zcrit}
    \Zcrithp \simeq \epsilon_{\rm{crit}}(\tau_s)\frac{\Hpsiturb}{H},
\end{equation}
where 
\begin{equation}\label{eq:Hpsiturb}
    \Hpsiturb \equiv\sqrt{\Hpsi^2+\Hpturb^2}.
\end{equation}

\noindent
where $\Hpsi = h_{\eta}\eta r$,  and $\Hpturb/H = \sqrt{\alphaDz/(\alphaDz+\tau_s)}$. The former represents the contribution of the SI-driven turbulence to the particle scale height, whereas the latter represents that of externally driven turbulence. We adopt the same value for $h_{\eta}$ as in \citetalias{Li21}, which is  $\simeq 0.2$. This value is an approximation for the particle scale height within the range of $\tau_s$ (and in units of $\eta r$) as measured from their SI simulations. As a result, $\Hpsi/H \simeq \Pi/5$. To obtain $\Hpturb$ for each $\alphaD$, we use $\alphaDz$ in Table \ref{table:gasinfo}.  

As can be seen from Figure \ref{fig:tauZmap}, the critical curve (sky-blue, Equation \ref{eq:LY21Zcrit}) and the data are not consistent at all; we find systematically lower critical $Z$ values than those calculated from Equation (\ref{eq:LY21Zcrit}), implying that Equation (\ref{eq:Hpsiturb}) does not accurately predict the actual scale height of particles in simulations.  The explanation for the inconsistency is given in Sections \ref{sec:results:feedbackhp} and \ref{sec:results:verticalprofile}, where we focus on the indirect impact of particle feedback on the particle scale height and the vertical profiles of particle density.

Since Equation (\ref{eq:LY21Zcrit}) does not match our simulation results, we attempt to provide a new fit to critical Z values. Assuming $\log{\Zcrit}$ is a function of both $\log{\tau_s}$ and $\log{\alphaD}$: 
\begin{equation}\label{eq:ourZcrit}
\begin{split}
    \log \Zcrit(\tau_s,\alphaD) &= A'(\log \alphaD)^2 + B'\log \tau_s\log \alphaD \\
                                &\quad + C'\log \tau_s + D'\log \alphaD,
\end{split}
\end{equation}
with conditions of 
\begin{equation}
     \alpha_{\rm{D,min}} = 10^{-3}\tau_s ~\text{and}~ 0.01 \leq \tau_s \leq 0.1, \nonumber
\end{equation}
where $A'=0.15$, $B'=-0.24$, $C'=-1.48$, and $D'=1.18$.  To find these coefficients, we performed a multivariate least squares fit, assuming (as we did for the $\epscrit$) that at a given $\tau_s$ and $\alphaD$, the critical $Z$ lies in the middle between adjacent empty (no collapse) and filled (collapse) circles. The resulting fits are shown as red curves in each panel of Figure \ref{fig:tauZmap}. The grey curves in each panel are random sample fits whose coefficients are randomly drawn from a multivariate normal distribution of $(A',B',C',D')$. As we did for $\epscrit$, we used a covariance matrix of $(A',B',C',D')$ to produce the normal distribution. We emphasize that Equation (\ref{eq:ourZcrit}) is valid only in the range of $\tau_s$ and $\alphaD$ provided above and should not be extrapolated beyond the range. This is because $\tau_s$ significantly affects particle dynamics; the simple form of $\Zcrit$ would not be able to encompass a wider range of $\tau_s$ than that we explore. In addition, we found that when $\alphaD \lesssim \alpha_{\rm{D,min}}$ at a given $\tau_s$, Equation (\ref{eq:ourZcrit}) has a turning point and ends up showing $\Zcrit$ increasing with decreasing $\alphaD$, which is unlikely.

It is also worth pointing out that we do not include $\Pi$ in Equation (\ref{eq:ourZcrit}). This is because while \citet{Sekiya2018} demonstrate that $Z/\Pi$ is the fundamental parameter combination (instead of $\Pi$ and $Z$ separately) for stratified SI, the role of $Z/\Pi$ in SI clumping when external turbulence is included has not yet been explored. 


\begin{figure*}
    \centering
    \includegraphics[width=\textwidth]{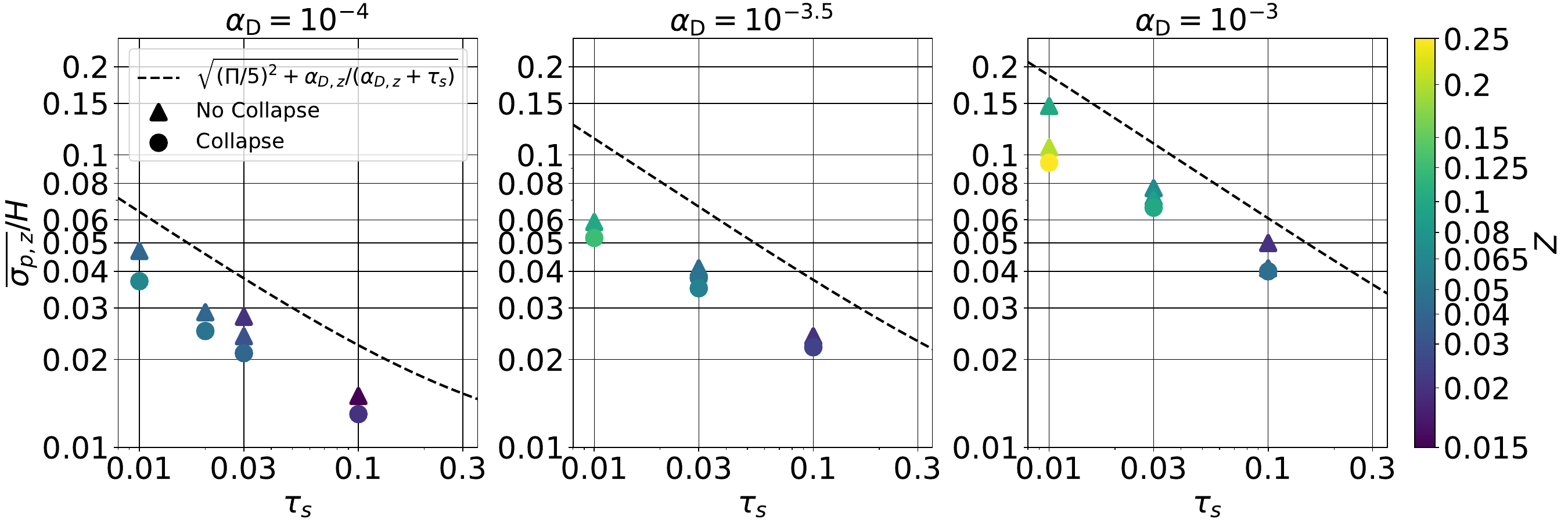}
    \caption{Similar to Figure \ref{fig:epsmap} but for the time-averaged standard deviation of particles' vertical positions (Equation \ref{eq:sigmapz}) in units of the gas scale height. The dashed black line in each panel denotes the prediction for the particle scale height described in Equation (\ref{eq:Hpsiturb}). As in Figure \ref{fig:epsmap}, we only present the simulations where we can do the time-average within a sufficiently long pre-clumping phase. The standard deviation measured in the simulations is always lower than the prediction. }
    \label{fig:Hpmap} 
\end{figure*}

\begin{figure*}
    \centering
    \includegraphics[width=\textwidth]{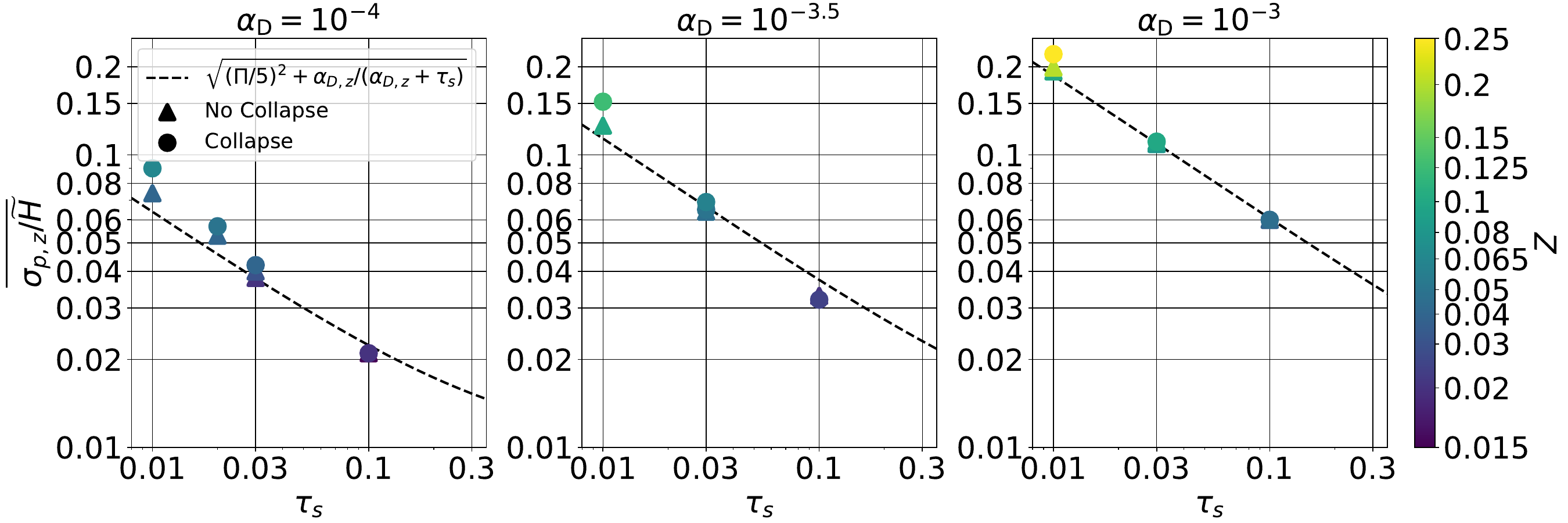}
    \caption{Similar to Figure \ref{fig:Hpmap} but for the time-averaged ratio of $\sigmapz$ to $\widetilde{H}$ (Equation \ref{eq:mixture_Hp}). We do not report results from simulations with a pre-clumping phase that is too short (as in Fig.~\ref{fig:Hpmap}). The figure shows that except for a few runs at $\tau_s=0.01$, this ratio is in excellent agreement with the dashed line (Equation \ref{eq:Hpsiturb}) for all $\tau_s$ at a given $\alphaD$.
    }
    \label{fig:hpbycsmap} 
\end{figure*}

 One of the interesting findings of \citetalias{Li21} is a sharp transition in the critical $Z$ value around $\tau_s \sim 0.015$. Although our parameter space data are too sparsely populated to examine this feature, the $\alphaD = 10^{-4}$ result implies that the sharp transition may disappear. As seen from the left panel of Figure \ref{fig:tauZmap}, $\Zcrit$ lies between 0.04 and 0.065 for $\tau_s = 0.01$ and between 0.04 and 0.05 for $\tau_s = 0.02$. This is a much shallower transition than was seen in \citetalias{Li21} in which the critical $Z$ is $\sim$ 0.016 and $\sim$ 0.007 for $\tau_s = 0.01$ and 0.02, respectively. To investigate this more, we plot particle density that is integrated in $z$ and averaged in $y$ versus $x$ and time (i.e., space-time plots) in Figure \ref{fig:Zdiscont}. The three panels correspond to different $\tau_s$ values, which are 0.01, 0.02, and 0.03, for $Z = 0.04$ and $\alphaD = 10^{-4}$. We did not turn on particle self-gravity during the time span considered in the figure. 

Figure \ref{fig:Zdiscont} reveals a stochastic evolution of the filaments; weak filaments are disrupted in all three cases, and only a few strong filaments survive in $\tau_s=0.02$ and $0.03$ cases. This may be due to the external turbulence, which can contribute to unevenly distributed filaments by providing additional diffusion. On the contrary, \citetalias{Li21} found that for $\tau_s = 0.01$ and $Z=0.0133$, filaments are so evenly spaced that they do not interact with each other, while those for $\tau_s = 0.02$ and $Z=0.01$ are less uniform, with the filaments merging with each other to form a few dense ones. Since we have only looked at the the jump in $\Zcrit$ for $\alphaD = 10^{-4}$, and even here, our data points around $\tau_s = 0.01$ are very sparse, further studies are needed to delve into this issue more. However, our results do imply that turbulence may disrupt the evenly spaced filaments that were seen in  \citetalias{Li21}, ultimately leading to a smoother transition in $\Zcrit$ between $\tau_s = 0.01$ and $\tau_s = 0.02$. It is also possible that the three-dimensional nature of the problem allows for this behavior, as the corresponding simulations in \citetalias{Li21} were all 2D.

\subsection{Particle Feedback and the Particle Scale Height}\label{sec:results:feedbackhp}
Massless particles settle toward the mid-plane while competing with turbulent stirring. This competition establishes a Gaussian distribution \citep{Dubrulle95} of the particle density with scale height $H_p$ that is related to the strength of turbulent stirring (i.e., $\alphaD$) and $\tau_s$ as in Equation~(\ref{eq:eqaz}).  However, for particles {\it with} mass, their mass (i.e, $Z$) can affect the vertical stirring as well by imposing mass loading on the gas, reducing $H_p$; the magnitude of this effect naturally depends on $Z$ (\citealt{Yang2017, Yang2018}, \citetalias{Li21}, \citealt{Xu22DustSettling}). Moreover, the particle feedback can alter the vertical profile of the particle density in other ways.  For example, \citet{Xu22DustSettling} carried out MHD simulations of the MRI in the low ionization limit and found that particle feedback enhances vertical settling by reducing the eddy correlation time and not by changing the vertical velocity of the gas. Furthermore, they found a non-Gaussian vertical particle density profile with a cusp around the mid-plane.  

In order to examine the effect of the particle feedback on the particle scale height ($H_p$), we calculate the standard deviation of the vertical particle position (which for a Gaussian profile would equal $H_p$):
\begin{equation}\label{eq:sigmapz}
    \sigmapz = \sqrt{\frac{1}{(N_{\rm{par}}-1)}\sum_{i=1}^{N_{\rm{par}}}(z_i-\langle z_i \rangle)^2},
\end{equation}
where $z_i$ is the vertical position of the $i$th particle and $\langle z_i \rangle$ is the mean vertical position. Figure \ref{fig:Hpmap} shows time-averaged $\sigmapz$ values from all runs in Figure \ref{fig:epsmap} as a function of $\tau_s$ and at each $\alphaD$. The color scale denotes $Z$ values. The circle and triangle markers show whether or not a run results in planetesimal formation via gravitational collapse (see Section \ref{sec:results:pltfrmtn} for details). The dashed line in each panel denotes the prediction for the Gaussian particle scale height when external turbulence is considered ($\Hpsiturb$; Equation \ref{eq:Hpsiturb}). The time-averaging is done during the pre-clumping phase to prevent the scale height from being skewed to the high density regions.


It is evident from Figure \ref{fig:Hpmap} that the measured $\sigmapz$ is always smaller than $\Hpsiturb$. Furthermore, at a given $\tau_s$ and $\alphaD$, larger $Z$ corresponds to smaller $\sigmapz$, indicating the particle feedback effect on the particle layer thickness. 

In order to understand this result in greater detail, we consider the effective scale height of a dust-gas mixture \citep{YangZhu2020MNRAS}
\begin{equation}\label{eq:mixture_Hp}
    \widetilde{H} = \frac{\widetilde{c_s}}{\Omega} = \frac{H}{\sqrt{1+\epsilon}}
\end{equation}
in which 
\begin{equation}\label{eq:mixture_cs}
    \widetilde{c_s} = \frac{c_s}{\sqrt{1+\epsilon}}
\end{equation}
is the effective sound speed of the mixture \citep{ShiChiang2013,LaibePrice2014,LinYoudin2017,ChenLin2018}; although the cited papers used $\rho_p/\rho_g$ to characterize particle density in Equation (\ref{eq:mixture_cs}), we decide to use $\epsilon$, which is the density ratio at the mid-plane, because our simulations are vertically stratified. As evident from the two equations above, increasing $\epsilon$ decreases the effective sound speed $(\widetilde{c_s}$) and the effective scale height  ($\widetilde{H}$). This can be interpreted as the effect of the mass loading of particles on the gas, which increases the mixture’s inertia but does not contribute to thermal pressure of the gas. With this in mind, we write the particle scale height in the presence of both external turbulence and the particle feedback as $H_p/\widetilde{H} = \Hpsiturb/H$ instead of $H_p = \Hpsiturb$. This results in:
\begin{equation}\label{eq:HpFeedbackTurb}
        H_p = \frac{H}{\sqrt{1+\epsilon}}\sqrt{\left(\frac{\Pi}{5}\right)^2+\frac{\alphaDz}{\alphaDz+\tau_s}}.
\end{equation}
In other words, the particle scale height is reduced by a factor of $\sqrt{1+\epsilon}$ from $\Hpsiturb$ in the presence of particle feedback. To compare this expected particle scale height with our results, we compute time-averages of $\sigmapz/\widetilde{H}=\sigmapz\sqrt{1+\epsilon}/H$ in the runs shown in Figure \ref{fig:Hpmap}, which should be close or equal to $\Hpsiturb/H$ (dashed lines in Figure \ref{fig:Hpmap}) if Equation (\ref{eq:HpFeedbackTurb}) accurately predicts $\sigmapz$.  Figure \ref{fig:hpbycsmap} shows the comparison between $\sigmapz/\widetilde{H}$ and $\Hpsiturb/H$. The figure clearly shows that most of the data points are on the dashed line (i.e., $\Hpsiturb/H$), which demonstrates that Equation (\ref{eq:HpFeedbackTurb}) predicts the scale height of particles from our simulations very accurately. This implies that the effect of the mass loading is likely the reason why $\sigmapz$ decreases with increasing $Z$ at a given $\tau_s$ and $\alphaD$ as found in Figure \ref{fig:Hpmap}. However, runs with $\tau_s = 0.01$ and high $Z$ values are still above the dashed lines, deviating from the prediction. We delve into these discrepancies in the next section.

\begin{figure*}
    \centering
    \includegraphics[width=\textwidth,]{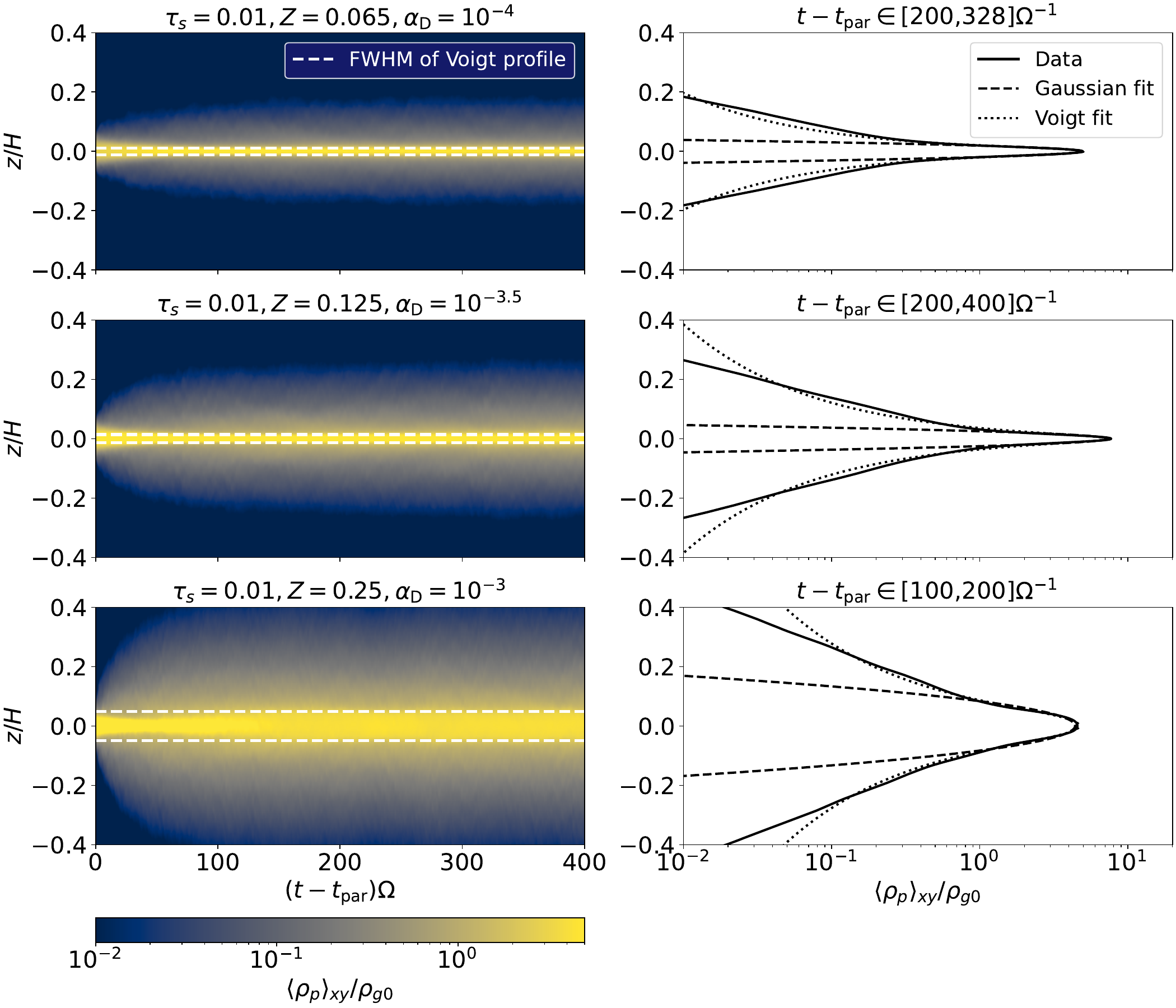}
    \caption{The space-time plots of particle density averaged in $x$ and $y$ ($\langle \rho_p \rangle_{xy}$) vs. $z$  and time  ($\it{left}$) and time-averaged vertical profiles of the particle density ($\it{right}$) for three runs at $\tau_s = 0.01$ that show $\sigmapz$ deviating from Equation (\ref{eq:HpFeedbackTurb}). Particle self-gravity is off during the time span considered here. From top to bottom, $\alphaD = 10^{-4}, 10^{-3.5}$, and $10^{-3}$. In the left panels, the horizontal dashed lines show the FWHM of the Voigt profiles shown in the right panels. The solid, dashed, and dotted lines in the right panels denote simulation data, Gaussian fit, and Voigt fit to the data, respectively. The title of each panel on the right shows the time interval over which each profile is averaged. Note that we only show the distributions between $z = \pm 0.4H$, while the vertical extents of the actual computational boxes are beyond this region. Particles form a thin, dense layer around the mid-plane and have a vertical density profile that deviates significantly from Gaussian. 
    }
    \label{fig:tau001rhop1dz} 
\end{figure*}

\begin{figure*}
    \centering
    \includegraphics[width=\textwidth,]{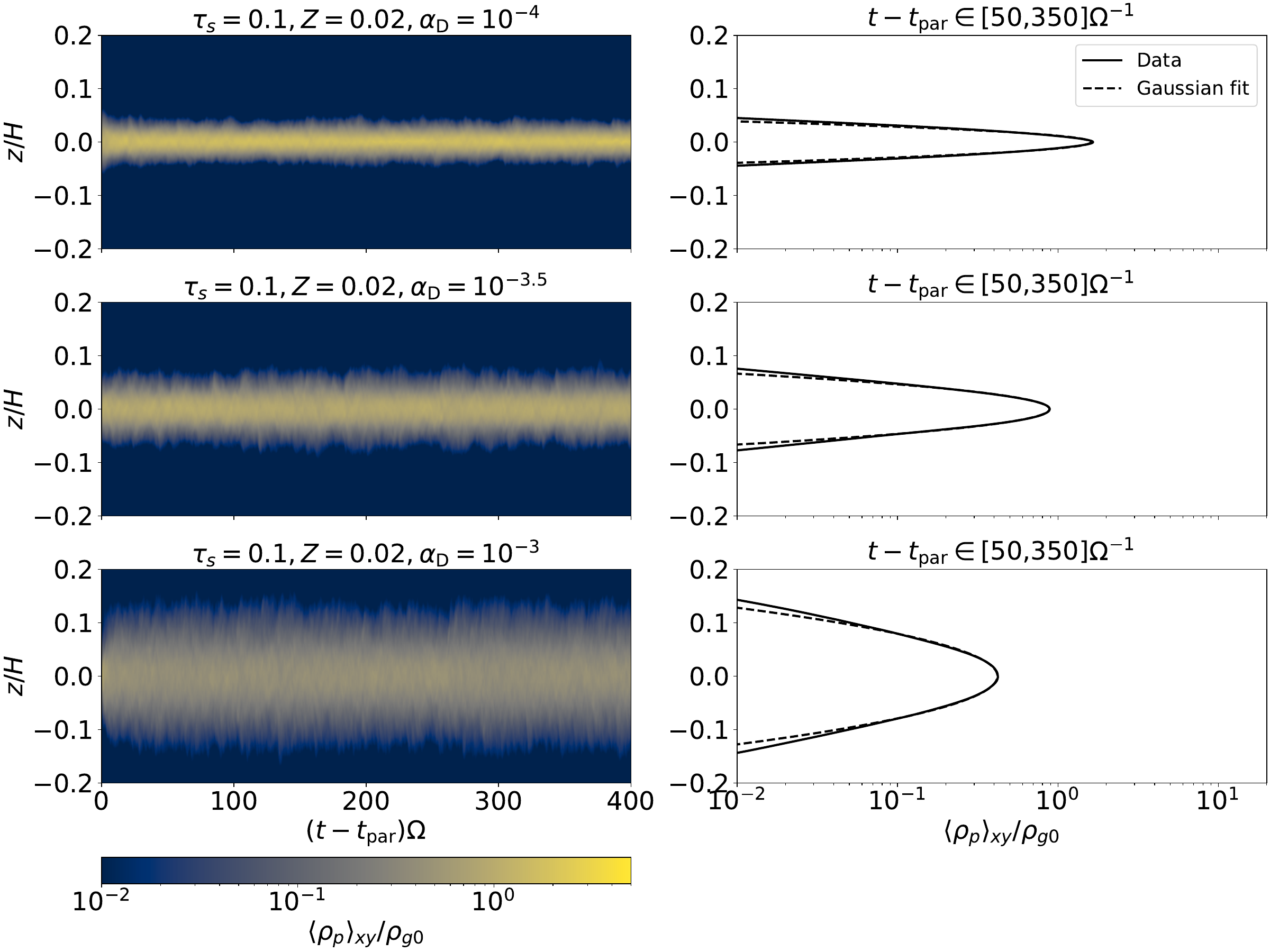}
    \caption{Similar to Figure \ref{fig:tau001rhop1dz} but for $\tau_s = 0.1, ~ Z=0.02$. Note that we show the distributions between $z = \pm 0.2H$ and do not show Voigt fits in the right panels.. Particle self-gravity is off during the time span presented here. Unlike $\tau_s$ = 0.01 cases, a thin, dense layer doesn't form, and the vertical profiles are closer to Gaussian.}
    \label{fig:tau01rhop1dz} 
\end{figure*}

\subsection{Vertical Distribution of Particle Density}\label{sec:results:verticalprofile}

As we just showed,  there are a few outliers at $\tau_s = 0.01$ that deviate from Equation (\ref{eq:HpFeedbackTurb}), namely Runs T10Z6.5A4, T1Z12.5A3.5, and T1Z25A3 all with $Z \gtrsim \Zcrit$. In what follows, we examine vertical profiles of the particle density in these simulations to explain the discrepancy with the prediction. The profiles are calculated with particle self-gravity turned off. 

In Figure \ref{fig:tau001rhop1dz}, we show the space-time plots of the horizontally averaged particle density vs. $z$ and time (left) and the time-averaged vertical profiles (right) for the three simulations. In all panels, we zoom in to $z = \pm 0.4H$ from the mid-plane. In the right panels, we present the Gaussian (dashed) and Voigt (dotted) fits to the simulation data (solid).

First, it is evident from the left panels that the particles build up a very thin layer close to the mid-plane, with an additional extended distribution vertically away from this layer. The vertical extent of both the thin layer and the extended distribution increases with increasing $\alphaD$. This is not surprising as we know that stronger vertical diffusion leads to a thicker layer.

Second, the plots on the right clearly reveal non-Gaussian particle density profiles. The profiles have a cusp near the mid-plane and extended wings on the outskirts of the cusp (see also \citealt{Xu22DustSettling} as they see a similar shape to the density distribution). The Gaussian fit (dashed) in each panel matches the data (solid) only very close to the mid-plane, whereas the Voigt fit (dotted) very approximately traces the data up to larger height (i.e., $\sim \pm 0.1H$ to $\sim \pm 0.2H$). The cusp develops due to the particle-loading on the gas that increases the inertia of the particle-gas fluid; this leads to the reduction of the vertical velocities of the gas at the mid-plane, with particles being more easily settled to form thin layers. Indeed, we found that $\sqrt{\langle u'_z{^2}\rangle}/c_s$ at the mid-plane is only $\sim 30\%$  of the same quantity averaged over the entire volume in the three simulations. On the other hand, some particles are still diffused away from the mid-plane and produce the extended region described above. We find that a Voigt profile, which has a Gaussian shape near the mid-plane and Lorentzian wings above and below the mid-plane, fits the simulation data much better than a Gaussian. Nonetheless, the data do deviate from the Voigt profile at sufficiently large $|z|$.

The discrepancy between $\sigmapz$ and Equation (\ref{eq:HpFeedbackTurb}) in the simulations with very large $Z$ is  simply the result of the density profile deviating significantly from a Gaussian.   Since Equations (\ref{eq:Hpsiturb}) and (\ref{eq:HpFeedbackTurb}) are based on a Gaussian profile, neither equation is appropriate for the scale height of particles that cause such strong feedback onto the gas.

We caution that a Voigt profile may not be an actual solution for the particle density distribution, and there is no physical motivation behind the fitting. Furthermore, \citet{LyraKuchnner2013} analytically derived a Gaussian particle density in the presence of particle feedback, which seems to contradict the non-Gaussian profiles in Figure \ref{fig:tau001rhop1dz}. This disagreement may stem from the fact that they assumed a constant diffusion coefficient. This is likely not true in our simulations since we find that vertical velocity of gas is reduced around the mid-plane (\citealt{Xu22DustTrap} also found that the gas velocity is reduced by particle mass-loading.). Future analytical and numerical studies are needed to examine the vertical profile of particle density in more depth. 

In an attempt to quantify the characteristic width of the thin layer, we calculate the Full Width at Half Maximum (FWHM) of the Voigt profile for each run presented in Figure \ref{fig:tau001rhop1dz}. The FWHM ($f_V$) is given by 

\begin{equation}\label{eq:Vogitfwhm}
    f_V = \frac{f_L}{2} + \sqrt{\frac{f_L^2}{4}+f_G^2},
\end{equation}
where $f_L$ and $f_G$ are FWHMs of Lorentzian and Gaussian profiles, respectively. The values of the $f_V$ for the three runs from top to bottom are $\sim 0.019H$, $\sim 0.032H$, and $\sim 0.097H$, respectively. We mark the FWHMs in the left panels as two horizontal dashed lines for each run; the FWHMs constrain the thickness of the thin particle layers very well. 

To demonstrate how the profiles change with larger $\tau_s$ values, we present plots similar to those in Figure \ref{fig:tau001rhop1dz}, but for $\tau_s = 0.1, ~ Z = 0.02$ in Figure \ref{fig:tau01rhop1dz}. Here, we zoom in further and present the distributions within $z = \pm 0.2H$, unlike than in Figure \ref{fig:tau001rhop1dz}. 

These runs show negligible discrepancy between $\sigmapz$ and Equation (\ref{eq:HpFeedbackTurb}) in Figure \ref{fig:hpbycsmap}.  Furthermore, as expected, the profiles are significantly changed compared to the $\tau_s = 0.01$ profiles. First, the vertical extent of the distributions is much narrower as a result of stronger settling of the $\tau_s = 0.1$ particles compared with the $\tau_s = 0.01$ particles. Second, thin particle layers that are separated from a more extended particle region do not form, and the time-averaged vertical profiles (right panels) do not show cusps at the mid-plane. This results from the $Z$ values being too small for particles to add considerable mass-loading on the gas. Third, as a result of this lack of the significant mass-loading, the density profiles are well approximated by a Gaussian but with reduced width according to Equation (\ref{eq:HpFeedbackTurb}). Overall, because of the absence of a cusp and Gaussian-like density profiles, $\sigmapz$ and Equation (\ref{eq:HpFeedbackTurb}) well matches each other than the simulations shown in Figure \ref{fig:tau001rhop1dz}.

In summary, the results in this subsection indicate two related considerations. First, due to mass-loading on the gas, the vertical particle density profile may still be Gaussian, but with a {\it reduced} width compared to the massless particle case. In this case, the criterion (Equation \ref{eq:LY21Zcrit}) {\it overestimates} critical $Z$ values. However, for even higher $Z$ values, the mass loading becomes so significant that the vertical particle density profile significantly deviates from Gaussian.  This deviation is particularly noticeable in simulations with $\tau_s = 0.01$, where $Z$ reaches such high levels that particles form a thin, dense layer at the mid-plane. Since an analytical expression for such a profile does not yet exist, we argue that caution is warranted when characterizing the width of particle layers under the influence of turbulence at large $Z$.  However, for more moderate $Z$ values Equation (\ref{eq:HpFeedbackTurb}) can be used to estimate particle scale height in the presence of both external turbulence and particle feedback.

\section{Discussion}\label{sec:Discussion}
This section is dedicated to providing a better understanding of the robustness and impact of our results. In Section \ref{sec:discussion:turb-ptl}, we consider the influence of turbulence on planetesimal formation, drawing comparisons with previous studies. We then discuss the potential implications of our findings on observations of protoplanetary disks in Section \ref{sec:discussion:observations}. Finally, we highlight potential limitations and uncertainties in our work in Section \ref{sec:discussion:caveats}.

\subsection{Does Turbulence Hinder or Help Planetesimal Formation?}\label{sec:discussion:turb-ptl}

While our results suggest that turbulence acts as a hindrance to planetesimal formation, there are a number of other possible routes to forming planetesimals in turbulent disks.  In this subsection, we discuss these other routes and the connection with this work.

\subsubsection{Self-consistently driven turbulence}
The results we present here stand in contrast to other work in which turbulence is included self-consistently (as opposed to forcing isotropic, incompressible turbulence as we do here) and gives rise to structures and behaviors that can concentrate particles. For example, the MRI is known to produce localized reductions in the radial pressure gradient (generally referred to as ``zonal flows"; see e.g., \citealt{Simon2014}). As radial drift slows in these regions (and can become trapped if the local pressure profile has a maximum), these zonal flows serve as natural sites for particle concentration and thus planetesimal formation as shown in, e.g., \cite{Johansen2007Rapidformation} and \cite{Xu22DustSettling}. Similarly,  \citet{SchaferVSIandSI} demonstrate that the VSI can also produce localized changes to the pressure gradient that then triggers the SI.\footnote{We note, however, that the simulations in \citet{SchaferVSIandSI} were 2D axisymmetric, and that 2D turbulence behaves differently from 3D (see \citealt {ALEXAKIS20181} for a review). Indeed, within the context of the SI, \citet{Sengupta_Umurhan23} recently demonstrated that 2D simulations exhibit notably different particle and gas dynamics compared to 3D calculations.}  On the contrary, there is no evidence of significant pressure variations to trap particles in our simulations, meaning that particle concentration is driven by the SI in this work.

At face value, these works suggest that turbulence can act to {\it help} planetesimal formation. More concretely, in the work by \citet{Xu22DustSettling}, the maximum density of particles surpassed $\rho_H$ for $\tau_s = 0.1$, $Z = 0.02$, even with $\alphaD \gtrsim 10^{-3}$. For comparison, our Run T10Z2A3 ($\tau_s = 0.1, Z = 0.02, \alphaD = 10^{-3}$) shows no indication of significant particle clumping at all (see the rightmost column of Figure \ref{fig:tau01-Z002-overview}). 

However, the potential discrepancy with our work can be elucidated by a more in-depth examination of $Z$.  In \cite{Xu22DustSettling}, the quoted $Z = 0.02$ value is a ``global" value (i.e., it was the average over the entire domain), and toward the pressure maximum (i.e., where $\Pi = 0$), $Z$ is enhanced.   While we cannot quantify the precise value of $Z$ at this location without more data, these considerations are in qualitative agreement with our results: the local value of $Z$ at the pressure maxima is (very possibly significantly) higher than the background value of $Z = 0.02$.

Such a direct comparison with \cite{Xu22DustSettling} should be treated with caution, however, as the particle concentrations in their work occur at the pressure maxima (i.e., where $\Pi = 0$); the SI does not operate in such regions. In fact, that they did not see particle concentration in regions outside of the pressure bump, where $Z$ is locally smaller and where $\Pi \ne 0$ aligns with our findings that planetesimal formation is hindered when  $\tau_s = 0.1, \ \alphaD = 10^{-3}$, and $Z \lesssim 0.04$. 

The key point here is that it is not the {\it turbulence} itself that is aiding in planetesimal formation (though, see arguments about the Turbulent Concentration mechanism below) but rather localized increases in $Z$ due to enhancements in the gas pressure. While this argument may appear to be one of semantics, it is important to distinguish between long-lived coherent structures, such as zonal flows, and random turbulent fluctuations, such as eddies, that are very short-lived by comparison.  Thus, while these pressure bumps are a side-effect of the turbulence, they are arguably distinct from the turbulence itself.  Our simulations provide a way to interpret these simulations by means of using $Z$ as a control parameter rather than one that changes with location based on the dynamics at work.

That all being said, \citet{Yang2018} demonstrate that particle concentration in a (at least somewhat) turbulent disk environment is possible even in the absence of such pressure bumps.  In particular, they find modest to strong particle concentration in an Ohmic dead zone (see e.g., \citealt{Gammie1996} for a description of the dead zone model) that results from an {\it anisotropy} in turbulent diffusion. That is, there was no pressure bump to trigger the SI but rather anisotropic turbulence.  A direct comparison with these results is difficult since our turbulence is forced isotropically, and thus we leave a study of the effect of anisotropic turbulence to future work.

Overall, our simulations should be treated as more controlled experiments; i.e., $Z$ is an input parameter and not something that arises from the simulation as a result of particle concentration. Furthermore, with the exception of \citet{SchaferVSIandSI}, all of the works involving turbulence driven by (magneto-)hydrodynamical instabilities in PPDs use $\tau_s = 0.1$, whereas our higher resolution simulations (i.e., more grid zones per $H$) allow us to resolve the SI for $\tau_s = 0.01$.\footnote{We use the results of \cite{Yang2017} as a guide. They saw filaments for $\tau_s = 0.01$ at a resolution equivalent to ours.} Thus, our work serves as an important counterpart to the numerous papers that include turbulence driven self-consistently and that (in most cases) are limited to larger $\tau_s$ by resolution requirements.

\subsubsection{Turbulent Concentration}

Beyond the processes we just described, planetesimal formation may be aided by turbulence through the turbulent concentration mechanism (hereafter, TC; \citealt{Cuzzi2001ApJ,Cuzzi2008ApJ,PhysRevE.95.033115,Hartlet_Cuzzi2020ApJ}). Specifically, particles with certain $\tau_s$ are preferentially concentrated by eddies whose eddy turnover time is comparable to $\tau_s$. In other words, there is an optimal $\tau_s$ value that makes the turbulent Stokes number at scale $\ell$ $\rm{St}_\ell \equiv \tau_s/\tau_{\rm{eddy},\ell} \sim 1$, where $\tau_{\rm{eddy},\ell}$, is the dimensionless eddy turnover time at scale $\ell$ ($\tau_{\rm{eddy},\ell} \equiv t_{\rm{eddy,\ell}}\Omega$, and $t_{\rm{eddy,\ell}}$ is the \textit{dimensional} turnover time). While we did not address the TC in this study, we now investigate whether TC might be present in our simulations by doing a simple scaling calculation. 

Assuming the forced turbulence follows the Kolmogorov relations, $u_\ell \propto \ell^{1/3}$ and $\ell \sim u_\ell t_{\rm{eddy},\ell} \propto t_{\rm{eddy},\ell}^{3/2}$, where $u_\ell$ and $t_{\rm{eddy},\ell}$ are a characteristic velocity and a turnover time of an eddy at scale $\ell$, respectively. Next, the forcing is done between $\sim 0.1H$ and $\sim 0.2H$ (see Appendix \ref{sec:appendix} for details), and we set $0.2H$ as the outer scale ($\mathcal{L}$) of the forced turbulence. The eddy turnover time for the outer scale is obtained from $\tau_{\rm{eddy},\mathcal{L}} \sim \alphaD/(\delta u/c_s)^2$ (Equation \ref{eq:eqaout}), which is $\sim$ 0.51, $\sim$ 0.44, and $\sim$ 0.37 for $\alphaD = 10^{-4}$, $10^{-3.5}$, and $10^{-3}$, respectively; $(\delta u/c_s)^2$ is computed without particles (see Table \ref{table:gasinfo}). For the purposes of this analysis, we set $\tau_{\rm{eddy},\mathcal{L}} = 0.44$. We thus obtain $\ell$ at which $\rm{St}_\ell\sim 1.0$ (i.e., $\tau_s = \tau_{\rm{eddy},\ell}$) where  TC becomes efficient:

\begin{equation}\label{eq:TC_scale_l}
    \ell = \mathcal{L}\left(\frac{\tau_s}{\tau_{\rm{eddy},\mathcal{L}}\rm{St}_{\ell}}\right)^{3/2} \sim 7\times 10^{-4}H\left(\frac{\tau_s}{0.01}\right)^{3/2}.
\end{equation}

Particles with $\tau_s = 0.01$ would be concentrated by eddies at scale $\ell \sim 7\times 10^{-4}H$, which is well below the width of a grid cell $\Delta = H/640 \sim 0.0016H$. The $\tau_s=0.03$ and $\tau_s=0.1$ particles on the other hand would be concentrated at $\ell \sim 0.004H$ (equating to $\ell/\Delta \sim 2$) and $\ell \sim 0.02H$ ($\ell/\Delta \sim 14$), respectively. These scales are above the grid scale (though in the case of $\tau_s = 0.03$, this is only marginally true).  However, a very approximate estimate for the dissipation scale (inferred from kinetic energy power spectra, see Appendix \ref{sec:appendix}) in our simulations is $0.01H$. The $\tau_s = 0.03$ particles would be concentrated on a scale less than this, whereas the $\tau_s = 0.1$ particles would be concentrated on a scale only twice that of the dissipation scale.

It is worth noting that while the most recent results suggest that $\rm{St}_\ell\approx 0.3$ is the most optimal value \citep{PhysRevE.95.033115,Hartlet_Cuzzi2020ApJ}, making it easier to resolve TC, there remains enough uncertainty in both this value and our approximate analysis that the absence of the TC should not be weighed too heavily. A much deeper dive into the TC as a possible mechanism for planetesimal formation is certainly required, but is beyond the scope of this paper.

In any case, our work clearly demonstrates that over a large parameter space of $\tau_s, Z$, and $\alphaD$ values, turbulence can significantly weaken the SI and prevent planetesimals from forming. This happens through turbulent diffusion counteracting vertical settling and/or the radial concentration of filaments within the disk plane.

\subsection{Implication for Observations}\label{sec:discussion:observations}
Several recent observations have quantified the scale height of millimeter-emitting dust in Class II disks, such as those surrounding HL Tau \citep{Pinte16} or Oph 163131 \citep{VillenaveHighSettling}. The findings from these studies generally suggest that the dust particles are well settled in the outer regions of the disks, with their scale heights being less than 1 au at a radial distance of 100 au. This behavior could imply the presence of very weak turbulence in the outer regions as indicated by the models used in the cited observations. However, those models neglected particle feedback. Thus, it is valuable to examine the potential implications for these observaiotns of particle feedback and the damping of turbulence.

\begin{figure*}
    \centering
    \includegraphics[width=\textwidth,]{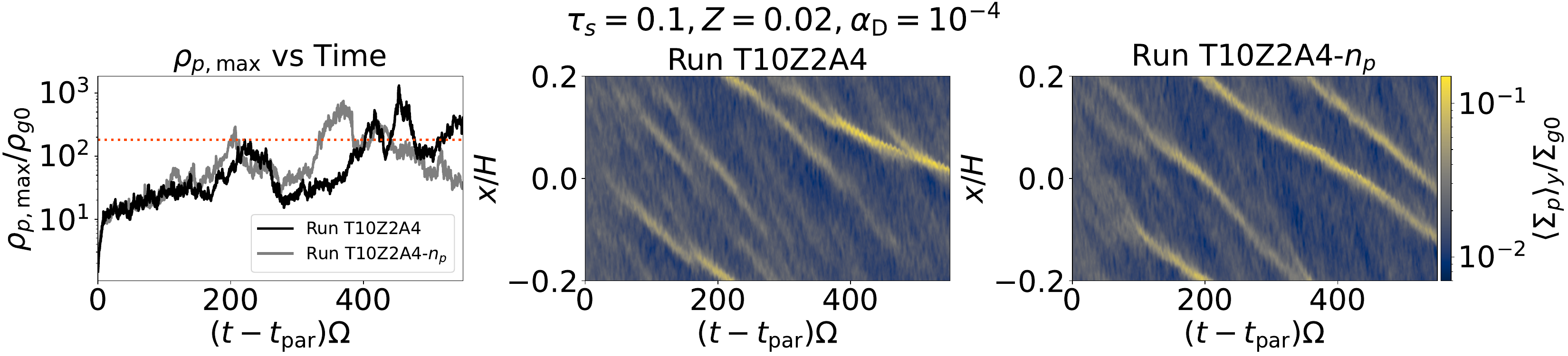}
    \caption{Comparison of two simulations with all parameters the same apart from the effective particle resolution ($n_{p,\rm{eff}}$). The left panel shows the time evolution of the maximum particle density, where the black and the grey curves are for the fiducial run (Run T10Z2A4, $n_{p,\rm{eff}} \sim 8$) and the run with the effective resolution of Run T10Z2A3 (Run T10Z2A4-$n_p$, $n_{p,\rm{eff}} \sim 8$), respectively. The orange horizontal line is the Hill density ($\rho_H$). The middle and  the right panels present the space-time plots of particle density averaged in $y$ and integrated over $z$ vs. $x$ and time for the two runs.  Particle self-gravity is disabled here. Despite slight differences in the maximum density and the radial concentration, the two runs with different $n_{p,\rm{eff}}$ are very similar. 
    }
    \label{fig:testnpar} 
\end{figure*}

\citet{VillenaveHighSettling} use a radiative transfer model to estimate the scale height of dust that radiates 1.3 mm continuum emission in the disk around Oph 163131. The resulting scale height is $\sim 0.5$ au at 100 au, while the scale height of the gas is estimated to be 9.7 $\pm$ 3.5 au from scattered-light data \citep{Wolff2021}. If we use the mean value for the gas scale height, this equates to $H_p/H \sim 0.05$. Assuming that the millimeter-emitting dust in their observation has $\tau_s = 0.01$ at 100 au and $\rho_p \ll \rho_g$,  $\alphaDz \sim 2.5 \times 10^{-5}$ (Equation \ref{eq:eqaz}). This $\alphaDz$ value is roughly consistent with the upper limit of $\sim 10^{-5}$ calculated in \cite{VillenaveHighSettling} by adopting the dust settling model of \citet{Fromang_Nelson2009}.  

However, as we discussed in Sections \ref{sec:results:feedbackhp} and \ref{sec:results:verticalprofile}, the mass of the particle can increase the inertia of the dust-gas mixture, resulting in a very thin particle layer even in the presence of stronger turbulence. For example, the dust thickness in our simulations, $\sigmapz/H$, spans from $\sim 0.04$ to $\sim 0.1$ for $\tau_s = 0.01$ depending on the values of $\alphaDz$ and $Z$ (see Figure \ref{fig:Hpmap}). The observed dust scale height in Oph~163131 equates to $H_p/H$ ranging from $\sim 0.04$ to $\sim 0.08$ if we account for the uncertainty in the gas scale estimation (i.e., $H = 9.7 \pm 3.5$ au), which {\it is} consistent with our results but only for $\alphaDz \gg 10^{-5}$.


To summarize, well-settled dust disks (i.e., small $H_p$) do not necessarily have very weak turbulence, especially if the dust-to-gas ratio is $\gg 1$. Thus, the level of turbulence inferred by ignoring feedback should be regarded as a lower limit.

\subsection{Caveats}\label{sec:discussion:caveats}
\subsubsection{Strength of Particle Self-Gravity}\label{sec:discussion:caveats:tildeG}
Throughout this paper, we fix $\tildeG$ (Equation \ref{eq:eqtildeG}) to 0.05 for all simulations we perform in order to compare our results to previous studies that assume (\citetalias{Li21}) or use the same value \citep{Gole20}. Moreover, our choice of the $\tildeG$ value should be viewed as a conservative way to establish the collapse threshold since planetesimal formation could be triggered in even weaker concentrations when a higher $\tilde{G}$ (or equivalently, a lower $Q$) is used (see Equation \ref{eq:Hill} or \citealt{Gerbig20}). In other words, critical $Z$ values will be lower in young massive disks or the outer regions of a Class II disk (e.g., $\tilde{G}$ $\sim$ 0.2 at $r = 45$ au based on the disk model in \citet{Carrera2021}). Therefore, our $\Zcrit$ in Section \ref{sec:results:pltfrmtn} is subject to change when different values of $\tildeG$ are used. 



\subsubsection{Number of Particles}\label{sec:discussion:caveats:numpar}
Every simulation considered in this work has the same average number of particles per grid cell (i.e., $n_p$ = 1). However, the effective particle resolution, which is the number of particles per $2\sigmapz$ (i.e., $n_{p,\rm{eff}} = n_p[L_z/2\sigmapz]$), varies with $\alphaD$ and the other parameters. For example, Run T10Z2A4 has $\sigmapz \sim 0.013$ and $n_{p,\rm{eff}} \sim 30$, whereas Run T10Z2A3 has $\sigmapz \sim 0.05$ and $n_{p,\rm{eff}} \sim 8$. In order to test how changing the effective particle resolution changes our results, we run Run T10Z2A4 again but with the same effective particle resolution of Run T10Z2A3, named T10Z2A4-$n_p$. The Run T10Z2A4-$n_p$ has $N_{\rm{par}} \sim 4.61 \times 10^{6}$, which is approximately a factor of 4 smaller than that of Run T10Z2A4. 

We compare the two models in Figure \ref{fig:testnpar}, which presents the maximum density of particles as a function of time in the left panel and the radial concentration over time in the middle and the right panels. In the left panel, Runs T10Z2A4 and T10Z2A4-$n_p$ are denoted as black and grey, respectively, and the Hill density at $\tilde{G} = 0.05$ is shown as the orange horizontal line. The particle self-gravity is turned off in the data shown here. From the maximum density plots, the two runs exhibit almost identical evolution until $t-t_{\rm{par}} \sim 200\Omega^{-1}$ after which they diverge. The time-averaged maximum densities from $t-t_{\rm{par}} = 200\Omega^{-1}$ to $500\Omega^{-1}$ are $\sim 138\rho_{g0}$ and $\sim 160\rho_{g0}$ for Runs T10Z2A4 and T10Z2A4-$n_p$, respectively. Given that the maximum density is inherently stochastic, the discrepancy does not seem to be significant. The radial concentration of particles of the two runs (middle and right panels) looks very similar as well. Although Run T10Z2A4-$n_p$ has one more filament at the end of the simulation, we do not believe this is a significant difference because the interaction between filaments is highly nonlinear, with the final number of dominant filaments and the maximum density values being uncertain to some degree. Overall, we conclude that the effective particle resolutions we employ is not likely to significantly affect our results. 

\begin{figure}
    \includegraphics[width=\columnwidth,]{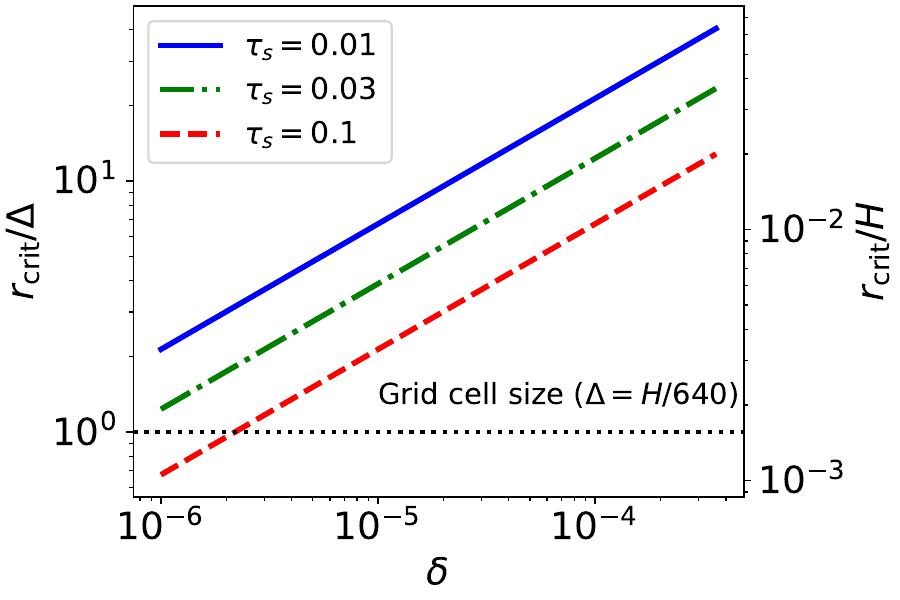}
    \caption{The critical radius of a particle cloud above which collapse occurs ($r_{\rm{crit}}$; Equation \ref{eq:rcrit}) as a function of $\delta$ for $\tau_s$ = 0.01 (blue), 0.03 (green), and 0.1 (red). The $y$-axes on the left and on the right show $r_{\rm{crit}}/\Delta$ and $r_{\rm{crit}}/H$, respectively, where $\Delta$ is the width of a grid cell. We assume that $\delta \approx \alphaDz$, which ranges from $10^{-6}$ to $\sim 4 \times 10^{-4}$. The larger number in this range corresponds to the value for $\alphaD = 10^{-3}$, while the smaller number is an assumed value based on the fact that particles become less diffusive around dense clumps. The black horizontal line denotes $\Delta$. Except for the case where $\delta$ is very low, the critical radius is larger than the cell size. 
    }\label{fig:figrcrit}
\end{figure}

\subsubsection{Numerical Resolution}\label{sec:discussion:caveats:resolution}

Other than $\tildeG$ and $n_p$, we also fix the grid resolution to $640/H$ in all simulations. However, it is possible that increasing this resolution would affect the $\epscrit$ and $\Zcrit$ curves. First considering 2D simulations, \citet{Yang2017} found that at $\tau_s=0.01$, the critical $Z$ above which the SI produces filaments lies between 0.02 and 0.04 at $640/H$ resolution, whereas it lies below 0.02 at $\gtrsim 1280/H$ resolution; these results suggest that increased resolution lowers the critical value. On the other hand, \citetalias{Li21} found that the critical $Z$ value above which strong clumping occurs (again, their definition of strong clumping requires the maximum particle density to exceed the Hill density) was 0.0133 (at $\tau_s = 0.01)$ for all resolutions up to $5120/H$ (though smaller boxes were used for higher resolution simulations). In their 3D simulations, \citet{Yang2017} found that for $\tau_s = 10^{-3}$ and $Z = 0.04$ at $160/H$ resolution, there was no sign of significant particle concentration or filament formation and $\rho_{p,\rm{max}} < 10\rho_{g0}$ (i.e., a critical $Z \gtrsim 0.04)$. However, at $320/H$ and $640/H$ resolutions, they observed a significant increase in concentration ($\rho_{p,\rm{max}} \sim 30\rho_{g0}$ and $\sim 200\rho_{g0}$ in the lower and the higher resolutions, respectively) and the formation of dense, persistent filaments (i.e., a critical $Z \lesssim 0.04)$. The differences in whether a resolution dependence for the critical $Z$ was observed may be a result of the different codes used (including other factors, such as boundary conditions). However, taken together, these results suggest that while resolution may affect the critical $Z$ values, these values will likely only be changed by a factor of order unity. Therefore, while more investigation is certainly needed to demonstrate if increasing grid resolution indeed affects $\epscrit$ and $\Zcrit$, we expect that it would produce at most a modest change in these critical values.

In addition to SI concentration, whether gravitational collapse of particles occurs depends on the numerical resolution. More specifically, numerical simulations should resolve the critical length $r_{\rm{crit}}$ derived in \citet{Klahr20} at which gravitational contraction balances internal (to a collapsing cloud of pebbles) turbulent diffusion to accurately capture the collapse and planetesimal formation.
\begin{equation}\label{eq:rcrit}
    r_{\rm{crit}} = \frac{1}{3}\sqrt{\frac{\delta}{\tau_s}}H.
\end{equation}
Here, $\delta$ is a dimensionless parameter for internal diffusion within a particle cloud. Particle clouds whose sizes are greater than $r_{\rm{crit}}$ are too massive to be held up by diffusion, and thus, they gravitationally collapse. Here, we examine whether or not the scale defined by $r_{\rm crit}$ is resolved in our simulations. Previous works measure the radial diffusion of particles, assuming they undergo a random walk in the radial direction, to quantify $\delta$ (see, e.g., \citealt{Baehr22,GerbigLi2023}). Instead of directly measuring the radial diffusion in our simulations, we let $\delta$ be a free parameter ranging from $10^{-6}$ to $4 \times 10^{-4}$ to cover $\delta$ values that likely result from the turbulence in our simulations. The lower limit is to account for the fact that denser regions are less diffusive \citep{GerbigLi2023}, and the upper limit corresponds to $\alphaDz$ for $\alphaD = 10^{-3}$. 

Figure \ref{fig:figrcrit} shows $r_{\rm{crit}}$ as a function of $\delta$ for three selected $\tau_s$ values, which are 0.01 (blue), 0.03 (green), and 0.1 (red). The first (left) and the second (right) vertical axes show $r_{\rm{crit}}$ in the unit of the size of a grid cell $(\Delta)$ and in the unit of $H$, respectively. We denote $\Delta$ as the black horizontal line. As can be seen, $r_{\rm{crit}}$ is generally larger than the grid cell size, meaning that our simulations should largely resolve gravitational collapse if a local particle clump is gravitationally unstable. For $\tau_s = 0.1$ (red) and very small $\delta$, the critical length becomes smaller than the cell size. Thus, it is possible that Runs T10Z1.5A4 and/or T10Z2A3.5 would produce collapsed regions if the numerical resolution was higher.  However, it is less likely that an increased resolution would change the results of the $\alphaD = 10^{-3}$ runs because the radial diffusion in these runs is much larger than $10^{-6}$. 

Overall, the $\epscrit$ and $\Zcrit$ we report should be viewed as upper limits since both SI concentration and gravitational collapse typically benefit from higher resolutions. Although our findings may be adjusted based on the caveats highlighted in this subsection, it is crucial to note that our simulations are 3D and incorporate essential physics, such as self-gravity of particles and external turbulence. Thus, our simulations represent a significant advancement in our quantification of the critical planetesimal formation curves.  

\section{Summary}\label{sec:summary}
In this paper we have presented results from stratified shearing box simulations in which gas and particles are aerodynamically coupled to each other. In order to study the effect of turbulence on the streaming instability (SI) and the formation of planetesimals, we include both the self-gravity of particles and externally driven incompressible turbulence. Our simulations explore a relatively broad range of parameter space, namely different dimensionless stopping times $\tau_s$, particle-to-gas surface density ratios $Z$, and forcing amplitudes $\alphaD$. We summarize our main results as follows:
\begin{enumerate}
    \item Incompressible turbulence can impede SI-driven concentration of particles via turbulent diffusion in two possible ways. First, this diffusion can prevent particles from settling, thereby preventing the mid-plane dust-to-gas density ratio $\epsilon$ from exceeding the critical value for filament formation. Second, even if the particles do settle and form a layer around mid-plane, the formation of filaments can be counterbalanced by turbulent diffusion acting in the plane of the disk (in addition to vertically). 

    \item  The critical $\epsilon$, at or above which planetesimal formation occurs, is $\epsilon \gtrsim 1$. This is a factor of a few larger than the corresponding values in the absence of externally driven turbulence, but is still of order unity.
    
    \item To balance the stronger diffusion associated with larger $\alphaD$, more total mass in the particles (i.e., through the $Z$ parameter) is needed. As such, the critical $Z$ values ($\Zcrit$) at or above which planetesimal formation occurs are much higher than those obtained without external turbulence. Quantitatively, when $\tau_s=0.01$, $\Zcrit$ $\sim 0.06$ and $\sim 0.2$ for $\alphaD = 10^{-4}$ and $10^{-3}$, respectively, whereas $Z \sim 0.02$ is sufficient for planetesimal formation in the absence of turbulence (e.g., \citetalias{Li21}).

    \item Due to particle feedback, the characteristic particle height in our simulations ($\sigmapz$) is always lower than the particle scale height with negligible feedback. This behavior is the direct result of enhanced particle mass loading on the gas.

    \item As a result of the strong influence of particle feedback on the dust scale height, observational measurements of the turbulent velocity should be regarded as a lower limit.  It is possible to have stronger turbulence {\it and} a small dust scale height if the dust-to-gas ratio is sufficiently large.
    
    \item  For sufficiently large $Z$, the vertical particle density profiles can be significantly modified from a Gaussian. For $\tau_s = 0.01$ and $Z \ge Z_{\rm crit}$, our simulations exhibit a cusp near the mid-plane resulting in a thin, dense layer of particles, and extended wings outside the layer, resembling a Voigt profile out to  $|z| \lesssim 0.2H$. 
\end{enumerate}
In closing, while there remain a number of uncertainties to be addressed in future work, our results demonstrate the crucial role that gas turbulence plays in limiting where in the disk and under what conditions planetesimals can form.

\section*{Acknowledgments} 
We thank Joanna Drazkowska, Konstantin Gerbig, Linn Eriksson, and Aleksandra Kuznetsova  for useful discussions related to this project. We thank Debanjan Sengupta and Orkan M. Umurhan for useful suggestions that improved the quality of this paper. J.L. and J.B.S acknowledge support from NASA under Emerging Worlds grant \# 80NSSC20K0702 and under the Theoretical and Computational Astrophysical Networks (TCAN) grant \# 80NSSC21K0497.   J.L. acknowledges support from NASA under the Future Investigators in NASA Earth and Space Science and Technology grant \# 80NSSC22K1322.
R.L. acknowledges support from the Heising-Simons Foundation 51 Pegasi b Fellowship.
CCY acknowledges the support from NASA via the Astrophysics Theory Program (grant \#80NSSC21K0141) and the Emerging Worlds program (grant \#80NSSC20K0347 and \#80NSSC23K0653).
The computations were performed using Stampede2 at the Texas Advanced Computing Center using XSEDE/ACCESS grant TG-AST120062.

\software{Julia \citep{bezanson2012julia},
          Athena \citep{Stone08,Stone2010,BaiStone10,Simon2016}}

\appendix
\counterwithin{figure}{section}
\section{Turbulence Forcing with Vector Potential}\label{sec:appendix}
In this appendix, we describe the vector potential driving used to force turbulence in the computational domain. In particular, we present the equations for the vector potential $\vecA$ and the method for handling the shearing-periodic boundary conditions in the radial direction. 

The vector potential is sinusoidal with phases that change over time; each component of the vector potential is defined as follows:
\begin{equation}\label{eq:vecpoentialA}
    \begin{split}
         A_x = \cos{[k_x(t)x+k_yy+\phi_{x1}(t)]}\cos{[k_zz+\phi_{x2}(t)]}, \\
         A_y= \cos{[k_x(t)x+k_yy+\phi_{y1}(t)]}\cos{[k_zz+\phi_{y2}(t)]}, \\
         A_z= \cos{[k_x(t)x+k_yy+\phi_{z1}(t)]}\cos{[k_zz+\phi_{z2}(t)]}. 
    \end{split}
\end{equation}
Here, $k_y = k_z = 2\pi/L_y = 2\pi/0.2H$ are the wavenumbers for the $y$ and $z$ directions, neither of which change with time. The radial (or, $x$) wavenumber changes with time and has a different form in order to make $\vecA$ and the resulting velocity perturbations consistent with background shear flow \citep{Hawley95}:
\begin{equation}\label{eq:kxt}
    k_x(t) = 2\pi n_x(t)/L_y + q\Omega k_y t,
\end{equation}
where $q \equiv -d\ln\Omega/d\ln R$; $q=3/2$ for a Keplerian disk. We employ a random number generator to change the phases (e.g., $\phi_{x1}(t)$) to guarantee that the vector potential is not correlated in time. The forcing occurs at intervals of $10^{-3}\Omega^{-1}$, and with each instance, the phase is assigned a new value.

In shearing box simulations, a computational domain is bordered radially by identical boxes, which shift azimuthally over time due to the shear flow; they are perfectly aligned at $t=0$. In this regard, by using $k_x(t)$ given above in our sinusoidal functions, we guarantee that $\vecA$ is in a co-moving frame with the shear flow. This construction guarantees that every component of the resulting velocity perturbations is continuous across the radial boundaries. In this setup, however, the second term on the right hand side of Equation \ref{eq:kxt} (and thus of course $k_x(t)$) increases over time without bound. Therefore, we repeatedly decrease $n_x(t)$ such that once $k_x(t)$ exceeds $2\pi/0.1H$, it lowers back to its initial value, so that $k_x(t) = 2\pi/0.2H$. The initial value and the upper bound of $k_x(t)$ are our choices, but are associated with long-wavelength modes (i.e., our turbulence is driven at large scales).

We numerically take the curl of $\vecA$ to generate each component of the velocity perturbations in real space, each of which is then multiplied by the gas density and added to the corresponding component of gas momentum (Equation \ref{eq:eq3}) with the desired amplitude of forcing (i.e., $\alphaD$). Since the perturbations should be located at grid cell centers to be consistent with how quantities are defined on the mesh, we shift $x,y$, and $z$ by the half-width of a grid cell. More specifically, for a given $A_i$ ($i=x, y$, or $z$),
\begin{equation}
    \begin{split}
        x \rightarrow x-0.5\Delta \ \text{if $i=y$ or $z$}, \\
        y \rightarrow y-0.5\Delta \ \text{if $i=x$ or $z$}, \\
        z \rightarrow z-0.5\Delta \ \text{if $i=x$ or $y$}, 
    \end{split}
\end{equation}
where $\Delta = H/640$ is the width of a grid cell. Finally, the velocity field injected via this method is incompressible (or, divergence-free) as taking the curl of the vector potential guarantees. To check the degree to which this incompressibility is maintained during the non-linear evolution of the system, we extract velocity components parallel to the wave-vector in Fourier space and carry out an inverse Fourier transform to produce them in real space; this gives us the curl-free velocity components. Then, we subtract them from the total velocity field to obtain the divergence-free components. In this way, we measure to what extent each component contributes to the total velocity field. We find that the divergence-free components account for $\sim 99\%$ of the total field, meaning that the forced turbulence is almost entirely incompressible.

\begin{figure*}
    \includegraphics[width=\columnwidth,]{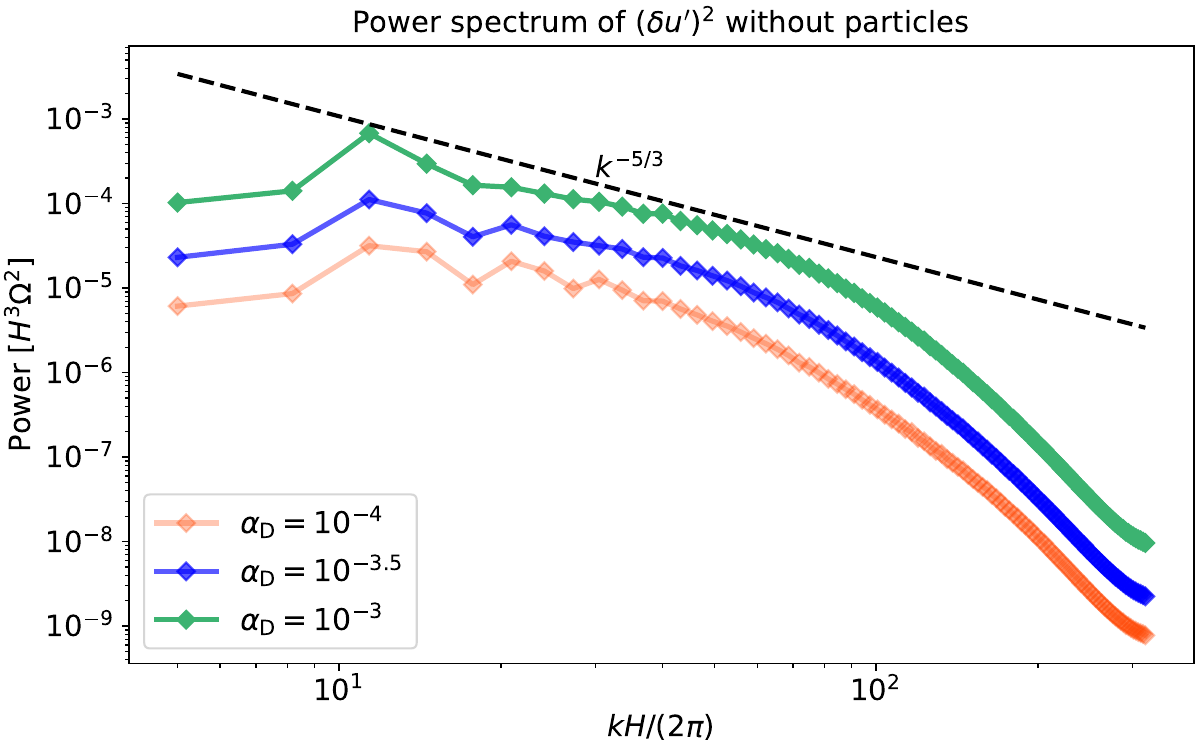}
    \caption{Time-averaged power spectra of turbulent velocity squared before initializing particles in simulations with $(L_x,L_y,L_z) = (0.4,0.2,0.8)H$. Three $\alphaD$ values are denoted by different colors as shown in the legend. The black dashed line represents Kolmogorov dependence $(\propto k^{-5/3}$). The power spectra have a peak at $kH/(2\pi) \sim 10$ followed by a power-law toward larger wavenumbers, which may be an indication of energy cascade. } Finally, at even larger $k$, the spectra become steeper due to numerical dissipation.
    \label{fig:powerspectrum}
\end{figure*}

Figure \ref{fig:powerspectrum} presents the time-averaged power spectra of the squared turbulent velocity (averaged over spherical shells of constant $k$ in Fourier space; see e.g., \citealt{Simon2012}) for simulations with three different $\alphaD$ and $L_x=0.4H,~L_y=0.2H,~L_z=0.8H$ and without particles. This is to demonstrate the proper implementation of forcing in our simulations. The color-coding denotes $\alphaD$ values. Additionally, the black dashed line shows a -5/3 power-law relation, which is a characteristic of Kolmogorov turbulence. The spectra are truncated at $kH/(2\pi) = 320$, which corresponds to the Nyquist scale $(2\Delta)$. We summarize key observations from the figure as follows:  

\begin{enumerate}
    \item Simulations with larger $\alphaD$ consistently show larger power across all considered scales.
    \item  The power spectra exhibit common features of turbulence in numerical simulations. That is to say, they peak at $kH/(2\pi) \sim 10$, which is a scale of $\sim 0.1H$. At $k$ larger than where the peak occurs, there is an indication of a (small) inertial range where a cascade of energy is likely taking place.  At even larger wavenumbers (or, smaller scales), the power-law is broken, and the slope becomes steeper due to numerical dissipation being dominant over inertia of the turbulence.

    \item Finally, the shape of the  power spectra's (small) inertial range is in rough agreement with a -5/3 power law, which suggests a Kolmogorov type behavior. However, we emphasize that our simulations, which are both rotating and stratified, will not strictly follow Kolmogorov phenomenology, and thus we cannot firmly conclude that the turbulence in our simulations is behaving as in the classical Kolmogorov picture. 

\end{enumerate}

\bibliography{main}{}

\begin{thebibliography}{}
\expandafter\ifx\csname natexlab\endcsname\relax\def\natexlab#1{#1}\fi
\providecommand{\url}[1]{\href{#1}{#1}}
\providecommand{\dodoi}[1]{doi:~\href{http://doi.org/#1}{\nolinkurl{#1}}}
\providecommand{\doeprint}[1]{\href{http://ascl.net/#1}{\nolinkurl{http://ascl.net/#1}}}
\providecommand{\doarXiv}[1]{\href{https://arxiv.org/abs/#1}{\nolinkurl{https://arxiv.org/abs/#1}}}

\bibitem[{Abod {et~al.}(2019)Abod, Simon, Li, Armitage, Youdin, \&
  Kretke}]{Abod2019}
Abod, C.~P., Simon, J.~B., Li, R., {et~al.} 2019, \apj, 883, 192,
  \dodoi{10.3847/1538-4357/ab40a3}

\bibitem[{{Adachi} {et~al.}(1976){Adachi}, {Hayashi}, \&
  {Nakazawa}}]{1976PThPh..56.1756A}
{Adachi}, I., {Hayashi}, C., \& {Nakazawa}, K. 1976, Progress of Theoretical
  Physics, 56, 1756, \dodoi{10.1143/PTP.56.1756}

\bibitem[{Alexakis \& Biferale(2018)}]{ALEXAKIS20181}
Alexakis, A., \& Biferale, L. 2018, Physics Reports, 767-769, 1,
  \dodoi{https://doi.org/10.1016/j.physrep.2018.08.001}

\bibitem[{Baehr {et~al.}(2022)Baehr, Zhu, \& Yang}]{Baehr22}
Baehr, H., Zhu, Z., \& Yang, C.-C. 2022, \apj, 933, 100,
  \dodoi{10.3847/1538-4357/ac7228}

\bibitem[{{Bai} \& {Stone}(2010a)}]{BaiStone2010ApJ}
{Bai}, X.-N., \& {Stone}, J.~M. 2010a, \apj, 722, 1437,
  \dodoi{10.1088/0004-637X/722/2/1437}

\bibitem[{Bai \& Stone(2010b)}]{BaiStone10}
Bai, X.~N., \& Stone, J.~M. 2010b, \apjs, 190, 297,
  \dodoi{10.1088/0067-0049/190/2/297}

\bibitem[{Balbus \& Hawley(1991)}]{Balbus1991}
Balbus, S.~A., \& Hawley, J.~F. 1991, \apj, 376, 214

\bibitem[{Bezanson {et~al.}(2012)Bezanson, Karpinski, Shah, \&
  Edelman}]{bezanson2012julia}
Bezanson, J., Karpinski, S., Shah, V.~B., \& Edelman, A. 2012, arXiv preprint
  arXiv:1209.5145

\bibitem[{Birnstiel {et~al.}(2012)Birnstiel, Klahr, \& Ercolano}]{Birnstiel12}
Birnstiel, T., Klahr, H., \& Ercolano, B. 2012, A\&A, 539,
  \dodoi{10.1051/0004-6361/201118136}

\bibitem[{Blum \& Wurm(2008)}]{Blum08}
Blum, J., \& Wurm, G. 2008, ARA\&A, 46, 21,
  \dodoi{10.1146/annurev.astro.46.060407.145152}

\bibitem[{Carrera {et~al.}(2015)Carrera, Johansen, \& Davies}]{Carrera2015}
Carrera, D., Johansen, A., \& Davies, M.~B. 2015, A\&A, 579,
  \dodoi{10.1051/0004-6361/201425120}

\bibitem[{Carrera {et~al.}(2021)Carrera, Simon, Li, Kretke, \&
  Klahr}]{Carrera2021}
Carrera, D., Simon, J.~B., Li, R., Kretke, K.~A., \& Klahr, H. 2021, The
  Astronomical Journal, 161, 96, \dodoi{10.3847/1538-3881/abd4d9}

\bibitem[{{Chen} \& {Lin}(2018)}]{ChenLin2018}
{Chen}, J.-W., \& {Lin}, M.-K. 2018, \mnras, 478, 2737,
  \dodoi{10.1093/mnras/sty1166}

\bibitem[{Chen \& Lin(2020)}]{Chen20}
Chen, K., \& Lin, M.-K. 2020, \apj, 891, 132, \dodoi{10.3847/1538-4357/ab76ca}

\bibitem[{{Colella}(1990)}]{Colella1990}
{Colella}, P. 1990, Journal of Computational Physics, 87, 171,
  \dodoi{10.1016/0021-9991(90)90233-Q}

\bibitem[{{Colella} \& {Woodward}(1984)}]{Colella1984}
{Colella}, P., \& {Woodward}, P.~R. 1984, Journal of Computational Physics, 54,
  174, \dodoi{10.1016/0021-9991(84)90143-8}

\bibitem[{{Cuzzi} {et~al.}(2001){Cuzzi}, {Hogan}, {Paque}, \&
  {Dobrovolskis}}]{Cuzzi2001ApJ}
{Cuzzi}, J.~N., {Hogan}, R.~C., {Paque}, J.~M., \& {Dobrovolskis}, A.~R. 2001,
  \apj, 546, 496, \dodoi{10.1086/318233}

\bibitem[{{Cuzzi} {et~al.}(2008){Cuzzi}, {Hogan}, \& {Shariff}}]{Cuzzi2008ApJ}
{Cuzzi}, J.~N., {Hogan}, R.~C., \& {Shariff}, K. 2008, \apj, 687, 1432,
  \dodoi{10.1086/591239}

\bibitem[{{Dominik} \& {Dullemond}(2023)}]{dominik2023bouncing}
{Dominik}, C., \& {Dullemond}, C. 2023, arXiv e-prints, arXiv:2312.06000,
  \dodoi{10.48550/arXiv.2312.06000}

\bibitem[{Dubrulle {et~al.}(1995)Dubrulle, Morfill, \& Sterzik}]{Dubrulle95}
Dubrulle, B., Morfill, G., \& Sterzik, M. 1995, ICARUS, 114, 237

\bibitem[{Flaherty {et~al.}(2020)Flaherty, Hughes, Simon, Qi, Bai, Bulatek,
  Andrews, Wilner, \& Ágnes Kóspál}]{Flaherty2020}
Flaherty, K., Hughes, A.~M., Simon, J.~B., {et~al.} 2020, \apj, 895, 109,
  \dodoi{10.3847/1538-4357/ab8cc5}

\bibitem[{Flaherty {et~al.}(2018)Flaherty, Hughes, Teague, Simon, Andrews, \&
  Wilner}]{Flaherty2018}
Flaherty, K.~M., Hughes, A.~M., Teague, R., {et~al.} 2018, \apj, 856, 117,
  \dodoi{10.3847/1538-4357/aab615}

\bibitem[{Flaherty {et~al.}(2017)Flaherty, Hughes, Rose, Simon, Qi, Andrews,
  Ágnes Kóspál, Wilner, Chiang, Armitage, \& ning Bai}]{Flaherty2017}
Flaherty, K.~M., Hughes, A.~M., Rose, S.~C., {et~al.} 2017, \apj, 843, 150,
  \dodoi{10.3847/1538-4357/aa79f9}

\bibitem[{{Fromang} \& {Nelson}(2009)}]{Fromang_Nelson2009}
{Fromang}, S., \& {Nelson}, R.~P. 2009, \aap, 496, 597,
  \dodoi{10.1051/0004-6361/200811220}

\bibitem[{{Fromang} \& {Papaloizou}(2006)}]{Fromang2006}
{Fromang}, S., \& {Papaloizou}, J. 2006, \aap, 452, 751,
  \dodoi{10.1051/0004-6361:20054612}

\bibitem[{{Gammie}(1996)}]{Gammie1996}
{Gammie}, C.~F. 1996, \apj, 457, 355, \dodoi{10.1086/176735}

\bibitem[{{Gerbig} \& {Li}(2023)}]{GerbigLi2023}
{Gerbig}, K., \& {Li}, R. 2023, \apj, 949, 81, \dodoi{10.3847/1538-4357/acca1a}

\bibitem[{Gerbig {et~al.}(2020)Gerbig, Murray-Clay, Klahr, \& Baehr}]{Gerbig20}
Gerbig, K., Murray-Clay, R.~A., Klahr, H., \& Baehr, H. 2020, \apj, 895, 91,
  \dodoi{10.3847/1538-4357/ab8d37}

\bibitem[{Gole {et~al.}(2020)Gole, Simon, Li, Youdin, \& Armitage}]{Gole20}
Gole, D.~A., Simon, J.~B., Li, R., Youdin, A.~N., \& Armitage, P.~J. 2020,
  \apj, 904, 132, \dodoi{10.3847/1538-4357/abc334}

\bibitem[{Güttler {et~al.}(2010)Güttler, Blum, Zsom, Ormel, \&
  Dullemond}]{Guttler10}
Güttler, C., Blum, J., Zsom, A., Ormel, C.~W., \& Dullemond, C.~P. 2010, A\&A,
  513, \dodoi{10.1051/0004-6361/200912852}

\bibitem[{{Hartlep} \& {Cuzzi}(2020)}]{Hartlet_Cuzzi2020ApJ}
{Hartlep}, T., \& {Cuzzi}, J.~N. 2020, \apj, 892, 120,
  \dodoi{10.3847/1538-4357/ab76c3}

\bibitem[{Hartlep {et~al.}(2017)Hartlep, Cuzzi, \& Weston}]{PhysRevE.95.033115}
Hartlep, T., Cuzzi, J.~N., \& Weston, B. 2017, Phys. Rev. E, 95, 033115,
  \dodoi{10.1103/PhysRevE.95.033115}

\bibitem[{Hawley {et~al.}(1995)Hawley, Gammie, \& Balbus}]{Hawley95}
Hawley, J.~F., Gammie, C.~F., \& Balbus, S.~A. 1995, \apj, 440, 742,
  \dodoi{10.1086/175311}

\bibitem[{{Johansen} {et~al.}(2015){Johansen}, {Mac Low}, {Lacerda}, \&
  {Bizzarro}}]{Johansen2015SciA}
{Johansen}, A., {Mac Low}, M.-M., {Lacerda}, P., \& {Bizzarro}, M. 2015,
  Science Advances, 1, 1500109, \dodoi{10.1126/sciadv.1500109}

\bibitem[{Johansen {et~al.}(2007)Johansen, Oishi, Low, Klahr, Henning, \&
  Youdin}]{Johansen2007Rapidformation}
Johansen, A., Oishi, J.~S., Low, M. M.~M., {et~al.} 2007, Nature, 448, 1022,
  \dodoi{10.1038/nature06086}

\bibitem[{{Johansen} \& {Youdin}(2007)}]{Johansen2007SIturb}
{Johansen}, A., \& {Youdin}, A. 2007, \apj, 662, 627, \dodoi{10.1086/516730}

\bibitem[{Johansen {et~al.}(2009)Johansen, Youdin, \& Low}]{Johansen09b}
Johansen, A., Youdin, A., \& Low, M. M.~M. 2009, \apj, 704,
  \dodoi{10.1088/0004-637X/704/2/L75}

\bibitem[{Klahr \& Schreiber(2020)}]{Klahr20}
Klahr, H., \& Schreiber, A. 2020, \apj, 901, 54,
  \dodoi{10.3847/1538-4357/abac58}

\bibitem[{Klahr \& Schreiber(2021)}]{Klahr21}
---. 2021, \apj, 911, 9, \dodoi{10.3847/1538-4357/abca9b}

\bibitem[{{Laibe} \& {Price}(2014)}]{LaibePrice2014}
{Laibe}, G., \& {Price}, D.~J. 2014, \mnras, 440, 2136,
  \dodoi{10.1093/mnras/stu355}

\bibitem[{{Lesur} {et~al.}(2022){Lesur}, {Ercolano}, {Flock}, {Lin}, {Yang},
  {Barranco}, {Benitez-Llambay}, {Goodman}, {Johansen}, {Klahr}, {Laibe},
  {Lyra}, {Marcus}, {Nelson}, {Squire}, {Simon}, {Turner}, {Umurhan}, \&
  {Youdin}}]{Lesur2023PPVII}
{Lesur}, G., {Ercolano}, B., {Flock}, M., {et~al.} 2022, arXiv e-prints,
  arXiv:2203.09821, \dodoi{10.48550/arXiv.2203.09821}

\bibitem[{Li \& Youdin(2021)}]{Li21}
Li, R., \& Youdin, A.~N. 2021, \apj, 919, 107, \dodoi{10.3847/1538-4357/ac0e9f}

\bibitem[{{Lin} \& {Youdin}(2017)}]{LinYoudin2017}
{Lin}, M.-K., \& {Youdin}, A.~N. 2017, \apj, 849, 129,
  \dodoi{10.3847/1538-4357/aa92cd}

\bibitem[{{Lyra} \& {Kuchner}(2013)}]{LyraKuchnner2013}
{Lyra}, W., \& {Kuchner}, M. 2013, \nat, 499, 184, \dodoi{10.1038/nature12281}

\bibitem[{{Manara} {et~al.}(2018){Manara}, {Morbidelli}, \&
  {Guillot}}]{Manara18}
{Manara}, C.~F., {Morbidelli}, A., \& {Guillot}, T. 2018, \aap, 618, L3,
  \dodoi{10.1051/0004-6361/201834076}

\bibitem[{Masset(2000)}]{Masset2000}
Masset, F. 2000, FARGO: A fast eulerian transport algorithm for differentially
  rotating disks

\bibitem[{{Nakagawa} {et~al.}(1986){Nakagawa}, {Sekiya}, \&
  {Hayashi}}]{Nakagawa1986}
{Nakagawa}, Y., {Sekiya}, M., \& {Hayashi}, C. 1986, \icarus, 67, 375,
  \dodoi{10.1016/0019-1035(86)90121-1}

\bibitem[{Nelson {et~al.}(2013)Nelson, Gressel, \& Umurhan}]{Nelson2013}
Nelson, R.~P., Gressel, O., \& Umurhan, O.~M. 2013, Monthly Notices of the
  Royal Astronomical Society, 435, 2610, \dodoi{10.1093/mnras/stt1475}

\bibitem[{{Paneque-Carre{\~n}o} {et~al.}(2023){Paneque-Carre{\~n}o},
  {Izquierdo}, {Teague}, {Miotello}, {Bergin}, {Loomis}, \& {van
  Dishoeck}}]{Paneque-Carreno2023}
{Paneque-Carre{\~n}o}, T., {Izquierdo}, A.~F., {Teague}, R., {et~al.} 2023,
  arXiv e-prints, arXiv:2312.04618, \dodoi{10.48550/arXiv.2312.04618}

\bibitem[{Pinte {et~al.}(2016)Pinte, Dent, Ménard, Hales, Hill, Cortes, \&
  de~Gregorio-Monsalvo}]{Pinte16}
Pinte, C., Dent, W. R.~F., Ménard, F., {et~al.} 2016, \apj, 816, 25,
  \dodoi{10.3847/0004-637x/816/1/25}

\bibitem[{{Sch{\"a}fer} {et~al.}(2017){Sch{\"a}fer}, {Yang}, \&
  {Johansen}}]{Schafer2017}
{Sch{\"a}fer}, U., {Yang}, C.-C., \& {Johansen}, A. 2017, \aap, 597, A69,
  \dodoi{10.1051/0004-6361/201629561}

\bibitem[{Schäfer \& Johansen(2022)}]{SchaferVSIandSI}
Schäfer, U., \& Johansen, A. 2022, A\&A, 666,
  \dodoi{10.1051/0004-6361/202243655}

\bibitem[{{Segura-Cox} {et~al.}(2020){Segura-Cox}, {Schmiedeke}, {Pineda},
  {Stephens}, {Fern{\'a}ndez-L{\'o}pez}, {Looney}, {Caselli}, {Li}, {Mundy},
  {Kwon}, \& {Harris}}]{2020NatureIRS63}
{Segura-Cox}, D.~M., {Schmiedeke}, A., {Pineda}, J.~E., {et~al.} 2020, \nat,
  586, 228, \dodoi{10.1038/s41586-020-2779-6}

\bibitem[{Sekiya \& Onishi(2018)}]{Sekiya2018}
Sekiya, M., \& Onishi, I.~K. 2018, \apj, 860, 140,
  \dodoi{10.3847/1538-4357/aac4a7}

\bibitem[{{Sengupta} \& {Umurhan}(2023)}]{Sengupta_Umurhan23}
{Sengupta}, D., \& {Umurhan}, O.~M. 2023, \apj, 942, 74,
  \dodoi{10.3847/1538-4357/ac9411}

\bibitem[{Shakura \& Sunyaev(1973)}]{SS73}
Shakura, N.~I., \& Sunyaev, R.~A. 1973, A\&A, 24, 337

\bibitem[{{Shi} \& {Chiang}(2013)}]{ShiChiang2013}
{Shi}, J.-M., \& {Chiang}, E. 2013, \apj, 764, 20,
  \dodoi{10.1088/0004-637X/764/1/20}

\bibitem[{{Simon} \& {Armitage}(2014)}]{Simon2014}
{Simon}, J.~B., \& {Armitage}, P.~J. 2014, \apj, 784, 15,
  \dodoi{10.1088/0004-637X/784/1/15}

\bibitem[{Simon {et~al.}(2016)Simon, Armitage, Li, \& Youdin}]{Simon2016}
Simon, J.~B., Armitage, P.~J., Li, R., \& Youdin, A.~N. 2016, \apj, 822, 55,
  \dodoi{10.3847/0004-637x/822/1/55}

\bibitem[{{Simon} {et~al.}(2012){Simon}, {Beckwith}, \& {Armitage}}]{Simon2012}
{Simon}, J.~B., {Beckwith}, K., \& {Armitage}, P.~J. 2012, \mnras, 422, 2685,
  \dodoi{10.1111/j.1365-2966.2012.20835.x}

\bibitem[{{Simon} {et~al.}(2022){Simon}, {Blum}, {Birnstiel}, \&
  {Nesvorn{\'y}}}]{SimonCometsIII}
{Simon}, J.~B., {Blum}, J., {Birnstiel}, T., \& {Nesvorn{\'y}}, D. 2022, arXiv
  e-prints, arXiv:2212.04509, \dodoi{10.48550/arXiv.2212.04509}

\bibitem[{Stone \& Gardiner(2010)}]{Stone2010}
Stone, J.~M., \& Gardiner, T.~A. 2010, \apjs, 189, 142,
  \dodoi{10.1088/0067-0049/189/1/142}

\bibitem[{Stone {et~al.}(2008)Stone, Gardiner, Teuben, Hawley, \&
  Simon}]{Stone08}
Stone, J.~M., Gardiner, T.~A., Teuben, P., Hawley, J.~F., \& Simon, J.~B. 2008,
  \apj Supplement Series, 178, 137, \dodoi{10.1086/588755}

\bibitem[{{Toomre}(1964)}]{Toomre1964}
{Toomre}, A. 1964, \apj, 139, 1217, \dodoi{10.1086/147861}

\bibitem[{Toro(2006)}]{Toro06}
Toro, E.~F. 2006, Riemann solvers and numerical methods for fluid dynamics : a
  practical introduction (Springer), 680

\bibitem[{Umurhan {et~al.}(2020)Umurhan, Estrada, \& Cuzzi}]{Orkan20}
Umurhan, O.~M., Estrada, P.~R., \& Cuzzi, J.~N. 2020, \apj, 895, 4,
  \dodoi{10.3847/1538-4357/ab899d}

\bibitem[{{Villenave} {et~al.}(2022){Villenave}, {Stapelfeldt}, {Duch{\^e}ne},
  {M{\'e}nard}, {Lambrechts}, {Sierra}, {Flores}, {Dent}, {Wolff}, {Ribas},
  {Benisty}, {Cuello}, \& {Pinte}}]{VillenaveHighSettling}
{Villenave}, M., {Stapelfeldt}, K.~R., {Duch{\^e}ne}, G., {et~al.} 2022, \apj,
  930, 11, \dodoi{10.3847/1538-4357/ac5fae}

\bibitem[{{Villenave} {et~al.}(2023){Villenave}, {Podio}, {Duch{\^e}ne},
  {Stapelfeldt}, {Melis}, {Carrasco-Gonzalez}, {Le Gouellec}, {M{\'e}nard}, {de
  Simone}, {Chandler}, {Garufi}, {Pinte}, {Bianchi}, \&
  {Codella}}]{VillenaveModestSettling}
{Villenave}, M., {Podio}, L., {Duch{\^e}ne}, G., {et~al.} 2023, \apj, 946, 70,
  \dodoi{10.3847/1538-4357/acb92e}

\bibitem[{{Wolff} {et~al.}(2021){Wolff}, {Duch{\^e}ne}, {Stapelfeldt},
  {M{\'e}nard}, {Flores}, {Padgett}, {Pinte}, {Villenave}, {van der Plas}, \&
  {Perrin}}]{Wolff2021}
{Wolff}, S.~G., {Duch{\^e}ne}, G., {Stapelfeldt}, K.~R., {et~al.} 2021, \aj,
  161, 238, \dodoi{10.3847/1538-3881/abeb1d}

\bibitem[{{Xu} \& {Bai}(2022a)}]{Xu22DustSettling}
{Xu}, Z., \& {Bai}, X.-N. 2022a, \apj, 924, 3, \dodoi{10.3847/1538-4357/ac31a7}

\bibitem[{{Xu} \& {Bai}(2022b)}]{Xu22DustTrap}
---. 2022b, \apjl, 937, L4, \dodoi{10.3847/2041-8213/ac8dff}

\bibitem[{Yang {et~al.}(2017)Yang, Johansen, \& Carrera}]{Yang2017}
Yang, C.~C., Johansen, A., \& Carrera, D. 2017, A\&A, 606,
  \dodoi{10.1051/0004-6361/201630106}

\bibitem[{Yang {et~al.}(2018)Yang, Low, \& Johansen}]{Yang2018}
Yang, C.-C., Low, M.-M.~M., \& Johansen, A. 2018, \apj, 868, 27,
  \dodoi{10.3847/1538-4357/aae7d4}

\bibitem[{{Yang} \& {Zhu}(2020)}]{YangZhu2020MNRAS}
{Yang}, C.-C., \& {Zhu}, Z. 2020, \mnras, 491, 4702,
  \dodoi{10.1093/mnras/stz3232}

\bibitem[{{Yang} \& {Zhu}(2021)}]{YangZhu2021MNRAS}
---. 2021, \mnras, 508, 5538, \dodoi{10.1093/mnras/stab2959}

\bibitem[{Youdin \& Johansen(2007)}]{YJ07}
Youdin, A., \& Johansen, A. 2007, \apj, 662, 613

\bibitem[{Youdin \& Goodman(2005)}]{YG05}
Youdin, A.~N., \& Goodman, J. 2005, \apj, 620, 459, \dodoi{10.1086/426895}

\bibitem[{{Youdin} \& {Kenyon}(2013)}]{2013pss3.book....1Y}
{Youdin}, A.~N., \& {Kenyon}, S.~J. 2013, in Planets, Stars and Stellar
  Systems. Volume 3: Solar and Stellar Planetary Systems, ed. T.~D. {Oswalt},
  L.~M. {French}, \& P.~{Kalas}, 1, \dodoi{10.1007/978-94-007-5606-9_1}

\bibitem[{Youdin \& Lithwick(2007)}]{Youdin_Lithwick2007}
Youdin, A.~N., \& Lithwick, Y. 2007, Icarus, 192, 588,
  \dodoi{10.1016/j.icarus.2007.07.012}

\bibitem[{Zsom {et~al.}(2010)Zsom, Ormel, Güttler, Blum, \&
  Dullemond}]{Zsom10}
Zsom, A., Ormel, C.~W., Güttler, C., Blum, J., \& Dullemond, C.~P. 2010, A\&A,
  513, \dodoi{10.1051/0004-6361/200912976}

\end{thebibliography}
\bibliographystyle{aasjournal}



\end{document}